\documentclass[twocolumn]{svjour3} 
\smartqed  % flush right qed marks, e.g. at end of proof
\usepackage{graphicx}
\begin{document}

\title{N$^*$ Experiments and what they tell us about Strong QCD Physics}
\author{V. D. Burkert}
\institute{V.D. Burkert \\
Thomas Jefferson National Accelerator Facility\\
Newport News, Virginia 23606, USA\\
              \email{burkert@jlab.org}   
}

\date{Received: date / Accepted: date}
\date{\today}

\maketitle

\begin{abstract}
I give an overview on experimental studies of the spectrum and the structure of the excited states of the 
nucleon and what we can learn about their internal structure. One focus is on the efforts to obtain a 
more complete picture of the light-quark baryon excitation spectrum employing electromagnetic beams that 
will allow us to draw some conclusions on the symmetries underlying the spectrum. For the higher mass 
excitations, the full employment of coupled channel approaches is essential when searching for new excited 
states in the large amounts of data already accumulated in different channels involving a variety
 of polarization observables. The other focus is on the study of transition form factors and helicity amplitudes and 
 their dependences on $Q^2$, especially on some of the more prominent resonances, especially    
$\Delta(1232)\frac{3}{2}^+$,
$N(1440)\frac{1}{2}^+$, and 
negative parity states 
$N(1535)\frac{1}{2}^-$, and 
$N(1675)\frac{5}{2}^-$.
These were obtained in pion and eta electroproduction experiments off proton targets and have already led to 
further insights in the active degrees-of-freedom as a function of the distance scale involved.  

\end{abstract}

\section{Excited baryon states in the history of the Universe}
\label{intro}

For this talk the organizers asked me to address what we learn about strong QCD (sQCD), i.e. QCD in the domain 
where perturbative methods fail in describing nucleon resonances transitions. 
Talking about a similar topic from the theory perspective, 
 Nathan Isgur said this in the concluding talk at $N*2000$:  "{\it I am convinced that completing this chapter in 
 the history of science will be one of the most interesting an fruitful areas of physics for at least the next thirty 
 years.}"  We are now 19 years into this 30 years prediction, and the physics of N*'s continues to go strong, 
 while many related issues remain to be explored.         

\begin{figure*}[h]
\centerline {
\includegraphics[width=4.5cm,height=7.0cm]{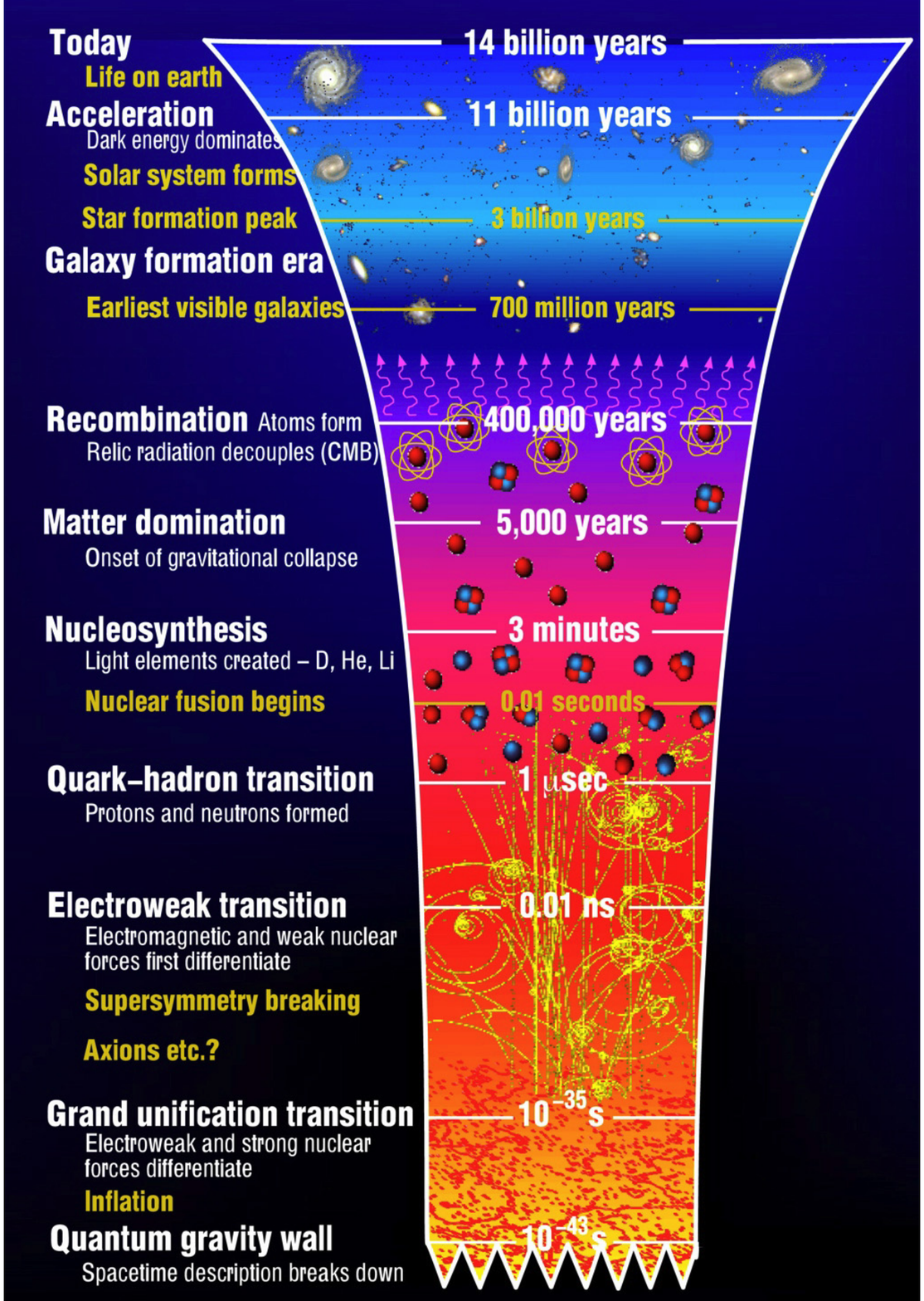} 
\hspace{0.5cm}\includegraphics[width=6.5cm,height=7.0cm]{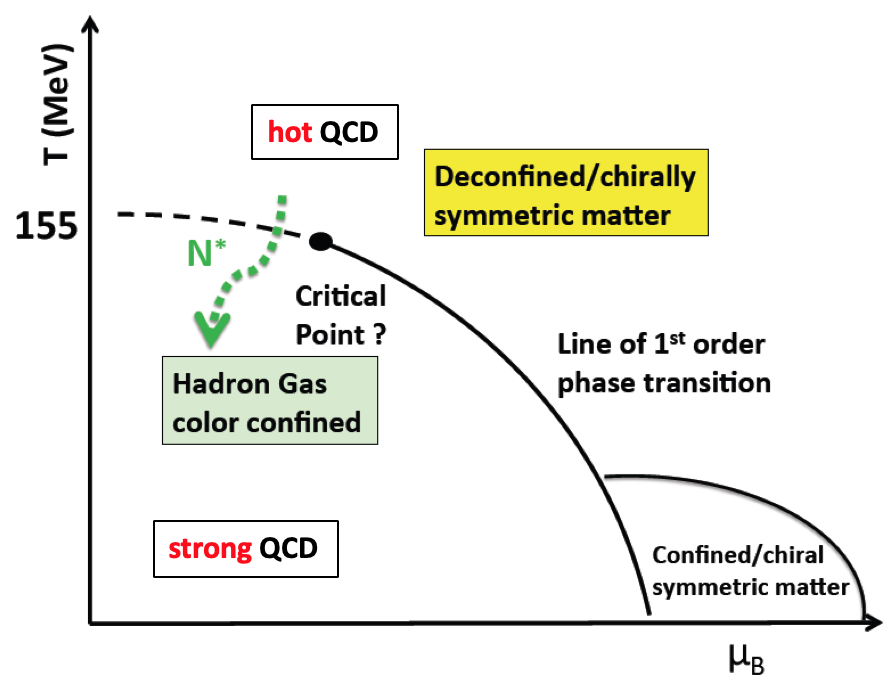}}
\caption{Left panel: The evolution of the Universe. 
The line denoted as {\it Quark-hadron transition}, is where protons and neutrons are formed. 
Existing electron accelerators as CEBAF, ELSA, and MAMI,
and colliders as BES III  have sufficient energy reach to access this region and study 
processes in isolation that occurred  during this transition in the microsecond old universe and resulted in the freeze 
out of baryons. Right panel: A generic phase diagram for the transition from the de-confined quark-gluon 
state to the confined hadron state.}
\label{universe}      
\end{figure*}

As we are trying to make progress in the complex world of physical sciences, we should not lose sight of 
what physics is all about: understanding the origin and the history of our universe, and the laws underlying
the observations. In this meeting we also address how excited states of the nucleon fit in to our understanding of the 
forces and the dynamics of matter in the history of the universe. On the internet we
find beautiful representations of the phases through which the universe evolved from the 
Big Bang (BB) to our times as shown in Fig.~\ref{universe}. 
There are some marked events that have been of particular significance during the early phases of the 
its history, such as the quark-gluon plasma of non-interacting colored quarks and gluons, and the 
forming of protons and neutrons. During this transition dramatic events occur - 
chiral symmetry is broken, quarks acquire mass dynamically, baryon 
resonances occur abundantly, and colored quarks and gluons are confined. This crossover 
process is governed by the excited hadrons, this is schematically 
shown in the generic QCD phase diagram in Fig.~\ref{universe}. In this process 
strong QCD (sQCD) is born as the process describing the interaction of colored quarks and gluons. These are the 
phenomena that we are exploring with electron and hadron accelerators - the full 
discovery of the baryon (and meson) spectrum, the role of chiral symmetry breaking and the 
generation of dynamical quark mass in confinement. While we can not recreate the exact condition in
the laboratory, with existing accelerators we can explore these processes in isolation. With electron machines
and high energy photon beams in the few GeV energy range we search for undiscovered excitations of 
nucleons and other baryons.  

As the Universe expands and cools down the coupling of quarks to the gluon field becomes stronger and 
quarks become more massive and form excited states in abundance.  This eventually gives 
way to the forming stable nucleons. In the heavy-quark sector It has been 
demonstrated~\cite{Bazavov:2014xya,Bazavov:2014yba} that the entire complement of 
excited states as predicted in the quark model~\cite{Capstick:1986bm} is needed to be included 
in these calculation to 
explain what is observed in "hot QCD" lattice calculations. Including only resonances from  
the Review of Particle Physics (RPP) is insufficient to explain the computations within hot QCD.
Similar projections have been made in the light-quark sector that we will discuss
 in the second part of this report. The close relationship of the baryon resonance spectrum and the 
 evolution of the early universe makes the experimental search for the "missing resonances" an even 
 more compelling experimental program. 
 \begin{figure*}[t]
\centering
\resizebox{2.0\columnwidth}{!}{\includegraphics{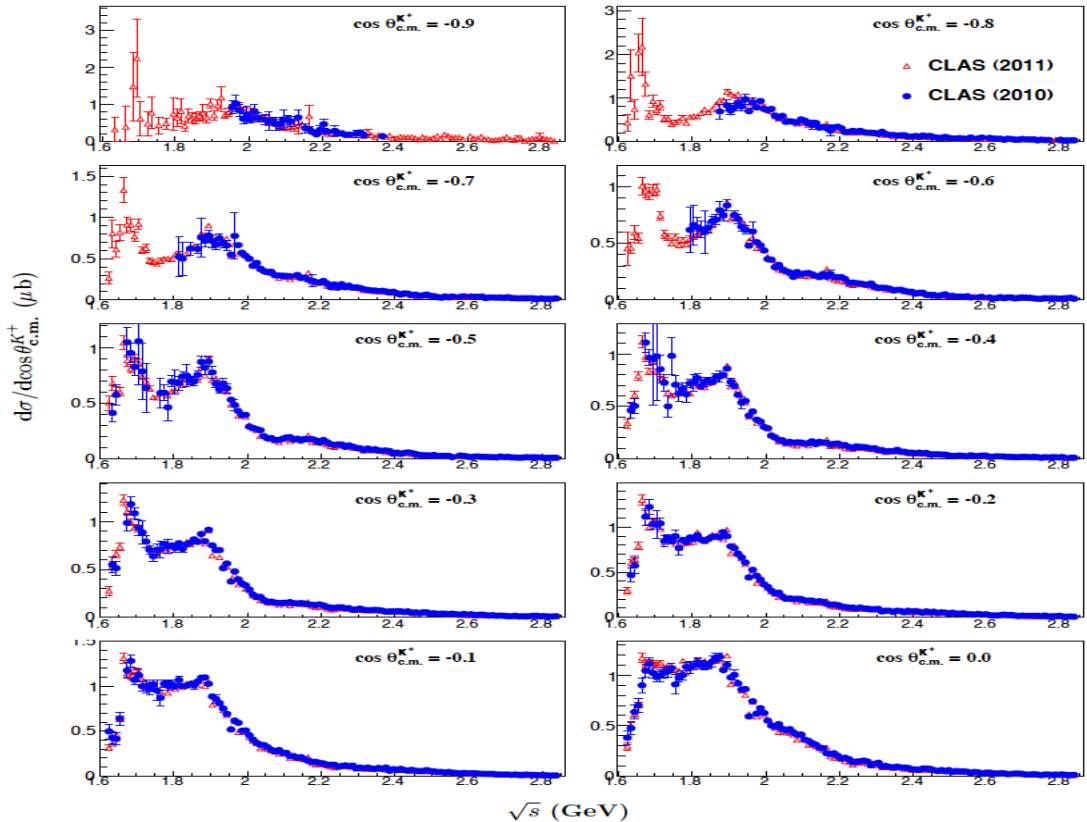}}
\caption{Invariant mass dependence of the $\gamma p \to K^+\Lambda$ differential cross section in the 
backward polar angle range. There are 3 structures visible that may indicate resonance excitations, 
at 1.7, 1.9, and 2.2 GeV. The blue full circles are based on the topology $K^+p\pi^-$, 
the red open triangles
are based on topology $K^+p$ or $K^+\pi^-$, which extended coverage towards lower W at backward angles
and allows better access to the resonant structure near threshold. }
\label{KLambda-crs}
\end{figure*}
 \section{Search for missing baryon states} 
Vigorous spectroscopy programs are currently underway at various particle accelerators 
in the quest for undiscovered excited mesons and baryons.   
 Experiments at electron machines such as CEBAF at JLAB in the US, ELSA at Bonn University in Germany, 
 and MAMI at the Johannes Gutenberg University at Mainz in Germany, 
 focus on the s-channel excitation of protons and neutrons to $\rm N^*$ and $\rm \Delta^*$ states.  

The excited states of the nucleon have been studied experimentally since the 1950's~\cite{Anderson:1952nw}. 
They contributed to the discovery of the quark model in 1964 by Gell-Mann and Zweig~\cite{GellMann1964,Zweig1964}, 
and were critical for the discovery of "color" degrees of freedom as first introduced by Greenberg~\cite{Greenberg:1964pe}. 
The 3-quark quark structure of baryons resulted in the prediction of a wealth of excited states  
with underlying spin-flavor and orbital symmetry of $SU(6) \otimes O(3)$. The predictions led to 
a broad experimental effort to search for these states. Of the many  
states predicted in the quark model, only a fraction have been observed, even today.  
Searches for the "missing" states and detailed studies of the resonance structure  
are now mostly carried out using electromagnetic probes and have been a major focus of  
hadron physics for the past two decades \cite{Burkert:2004sk}. A broad  
experimental effort has been underway for the past two decades, with measurements 
of  exclusive meson photoproduction and electroproduction reactions, including 
many polarization observables. Precision data and the development of multi-channel 
partial wave analysis procedures have resulted in the discovery of a series of excited states of the 
nucleon in the mass range of 1.9 to 2.2 GeV, others have been upgraded in their status of likelihood 
of existence as entered in the bi-annual Reviews of Particle Physics~\cite{Beringer:1900zz,Agashe:2014kda,Patrignani:2016xqp,Tanabashi:2018oca}.   
We will discuss some of these states in the following sections. 
\begin{figure*}[t]
\centering
\resizebox{2.0\columnwidth}{!}{\includegraphics{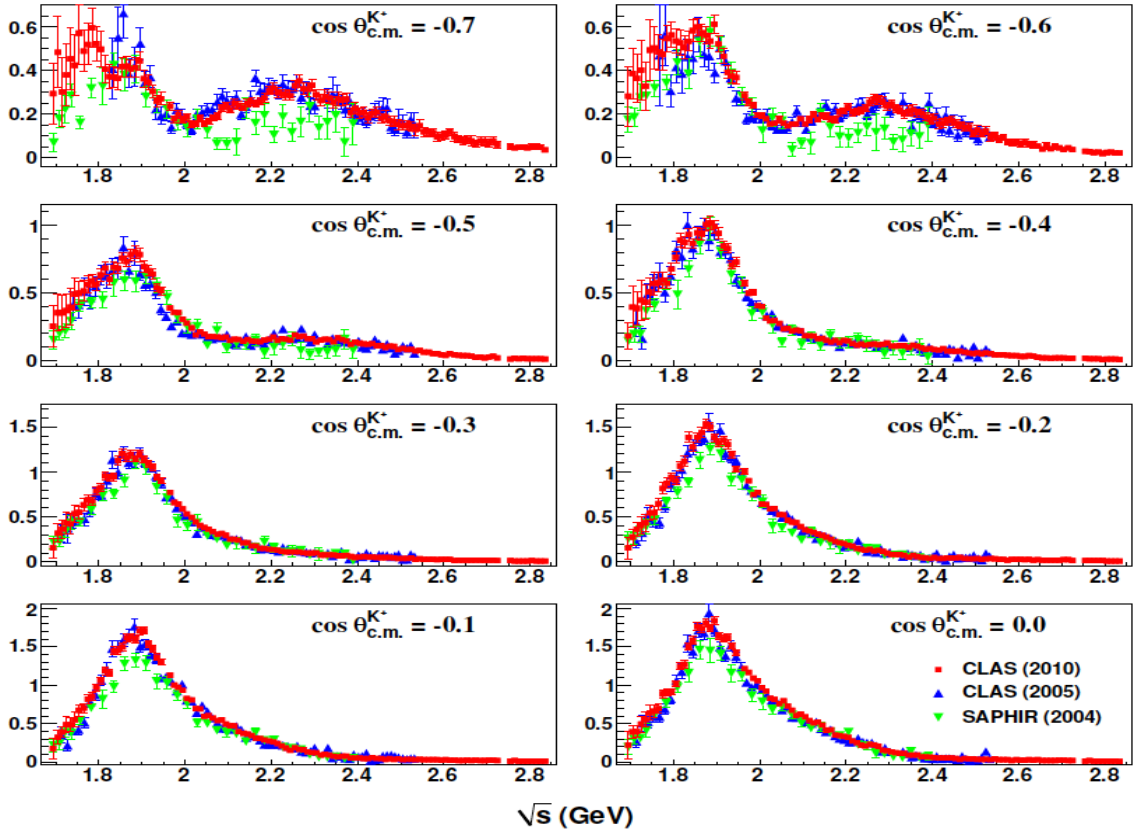}}
\caption{Invariant mass dependence of the $\gamma p \to K^+\Sigma^\circ$ differential cross section in the backward
polar angle range. }
\label{KSigma-crs}
\end{figure*}

Accounting for the complete excitation spectrum of the nucleon (protons and neutrons) 
and understanding the effective degrees of freedom is perhaps the most important and  
certainly the most challenging task of hadron physics. The experimental N* program currently 
focusses on the search for new excited states in the mass range above 2 GeV 
using energy-tagged photon beams in the few GeV range, and the study of the internal
structure of prominent resonances in meson electroproduction.  

A quantitative description of baryon spectroscopy and the structure of excited 
nucleons must eventually involve solving QCD for a complex strongly interacting 
multi-particle system. 
Recent advances in Lattice QCD led to predictions of the nucleon spectrum in QCD with 
dynamical quarks~\cite{Dudek:2012ag}, albeit with still large pion 
 masses of 396 MeV. Lattice prediction can therefore only be taken as indicative of the 
 quantum numbers of excited states and not of the masses of specific states. In parallel, 
 the development of dynamical coupled channel models is being pursued with new vigor. 
 The EBAC group at JLab as well as others have shown~\cite{Suzuki:2009nj} that dynamical effects can result 
 in significant mass shifts of the excited states. As a particularly striking result, a very large 
 shift was found for the Roper resonance pole mass to 1365 MeV downward from its bare 
 mass of 1736 MeV. This result has clarified the longstanding puzzle of the incorrect 
 mass ordering of $N(1440){1\over 2}^+$ and $N(1535){1\over 2}^-$ resonances in the constituent 
 quark model. Developments on the phenomenological side go hand in hand with a 
 world-wide experimental effort to produce high precision data in many different channel 
 as a basis for a determination of the light-quark baryon 
 resonance spectrum. On the example of experimental results from CLAS, the strong impact 
 of precise meson photoproduction data is discussed.  
 Several reviews have recently been published on this and related 
 subjects~\cite{Klempt:2009pi,Tiator:2011pw,Aznauryan:2011qj,Aznauryan:2012ba,Crede:2013sze} where
 many details can be found.

\section{Establishing the N* Spectrum}
\label{sec:1}
The complex structure of the light-quark (u \& d quarks) baryon excitation spectrum complicates the experimental 
search for individual states. As a result of the strong interaction, resonances are wide, often 200 MeV to 300 MeV, 
and are difficult to uniquely identify when only differential cross sections are measured. Most of the excited 
nucleon states listed in the Review 
of Particle Physics (RPP) prior to 2012 have been observed in elastic pion scattering 
$\pi N \to \pi N$. However there are important limitations in the sensitivity to the 
higher mass nucleon states that may have very small $\Gamma_{\pi N}$ decay widths, and   
the extraction of resonance contributions then becomes exceedingly difficult in this channel. 

Estimates for alternative decay channels have 
been made in quark model calculations\cite{Capstick:1993kb} for various channels.  This has
 led to a major experimental effort at Jefferson Lab, ELSA, GRAAL, and MAMI
to chart differential cross sections and polarization observables for a variety of meson
 photoproduction channels. At JLab with CLAS, many different final states have 
 been measured with high precision on the proton~\cite{Dugger:2005my,Dugger:2009pn,Mattione:2017fxc,Senderovich:2015lek,Williams:2009yj,Collins:2017sgu,Williams:2009aa,Williams:2009ab,Bradford:2006ba,Bradford:2005pt,Ho:2017kca,McCracken:2009ra,Paterson:2016vmc,Strauch:2015zob,Dey:2010hh,McNabb:2003nf,Golovatch:2018hjk}, many of them are now employed 
 in single- and in multi-channel analyses~\cite{Anisovich:2017pox}. Recently, the first measurements of open strangeness processes on neutron targets have been published~\cite{Compton:2017xkt,Ho:2018riy}.  

 \begin{figure}[t]
\centering
\resizebox{1.0\columnwidth}{!}{\includegraphics{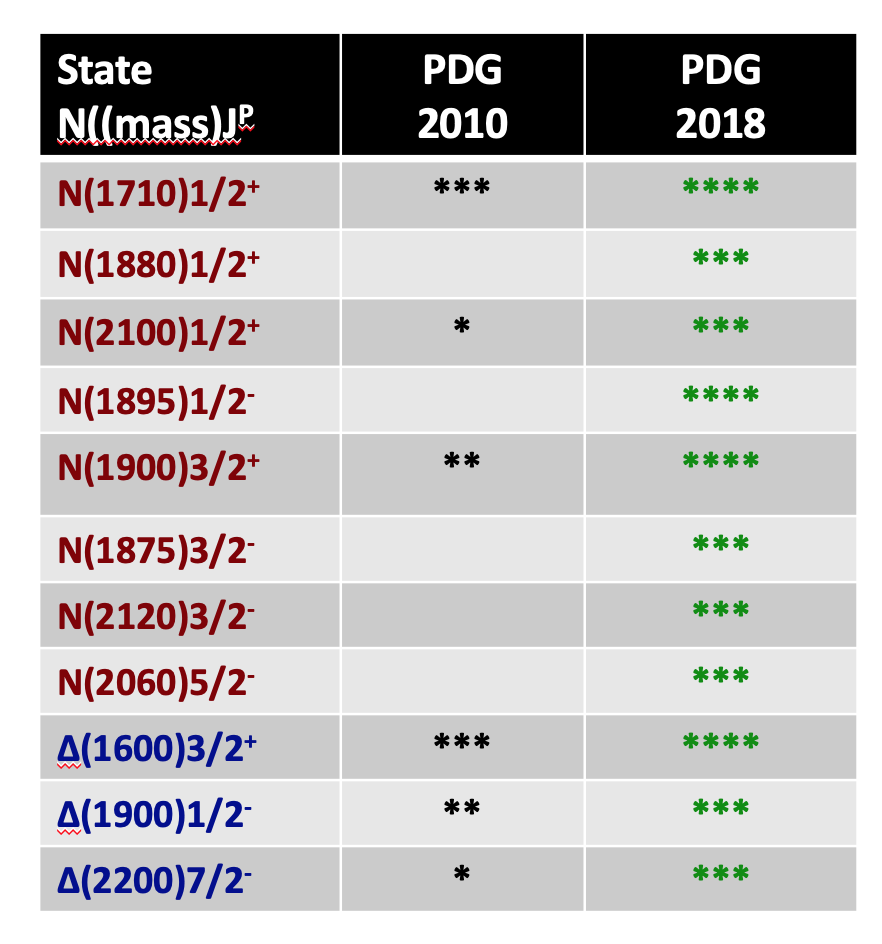}}
\caption{Evidence for 11 N and $\Delta$ states in RPP 2010 compared with RPP 2018~\cite{Tanabashi:2018oca}.}
\label{pdg2014}
\end{figure}
 
 %Moreover, 
 %some of these channels have also been explored in electroproduction reactions that add
 %another level of sensitivity to the search for new excited states~\cite{Mokeev}. 

\begin{figure}[ht]
\resizebox{1.0\columnwidth}{!}{\includegraphics{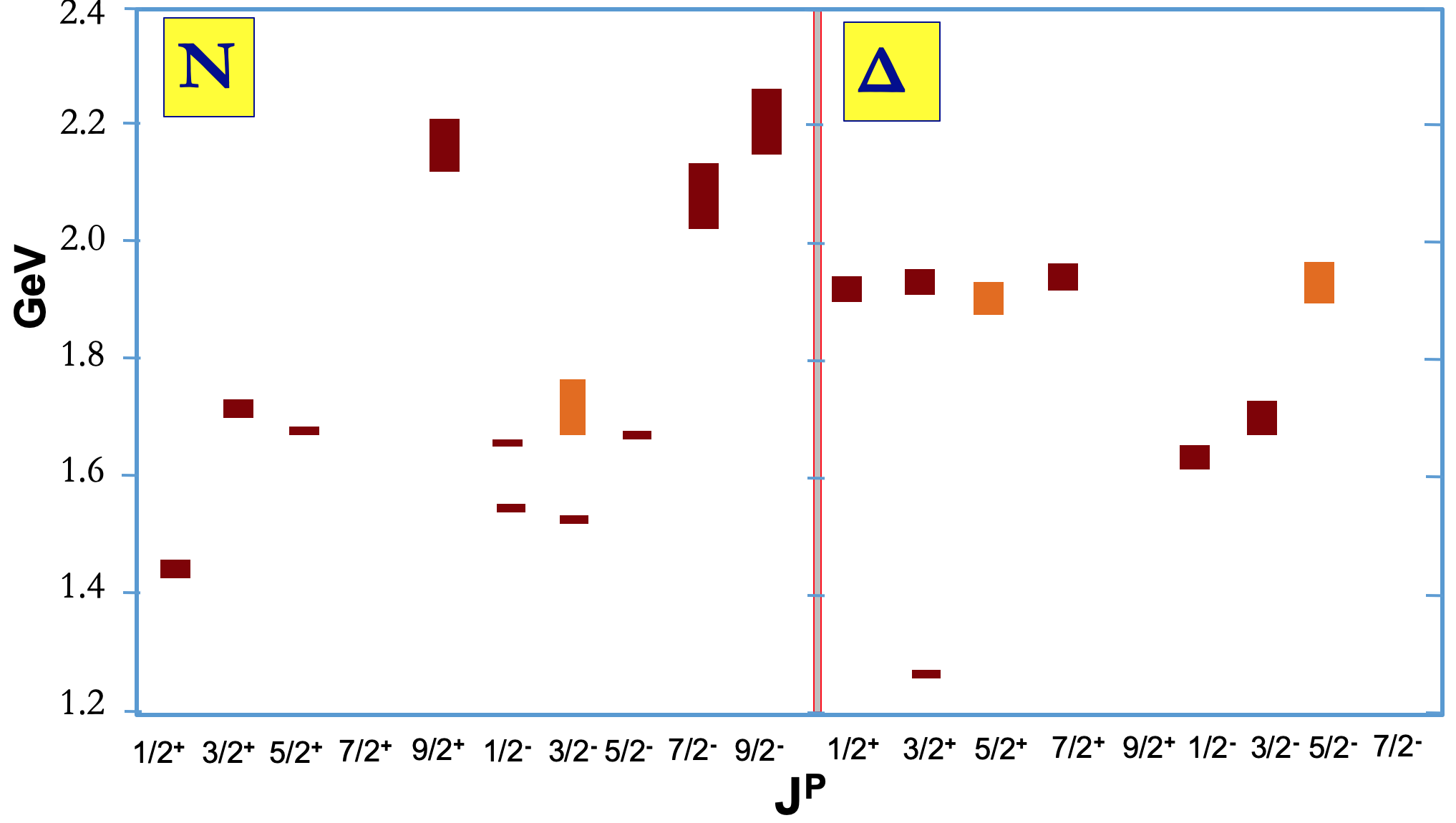}}
\resizebox{1.0\columnwidth}{!}{\includegraphics{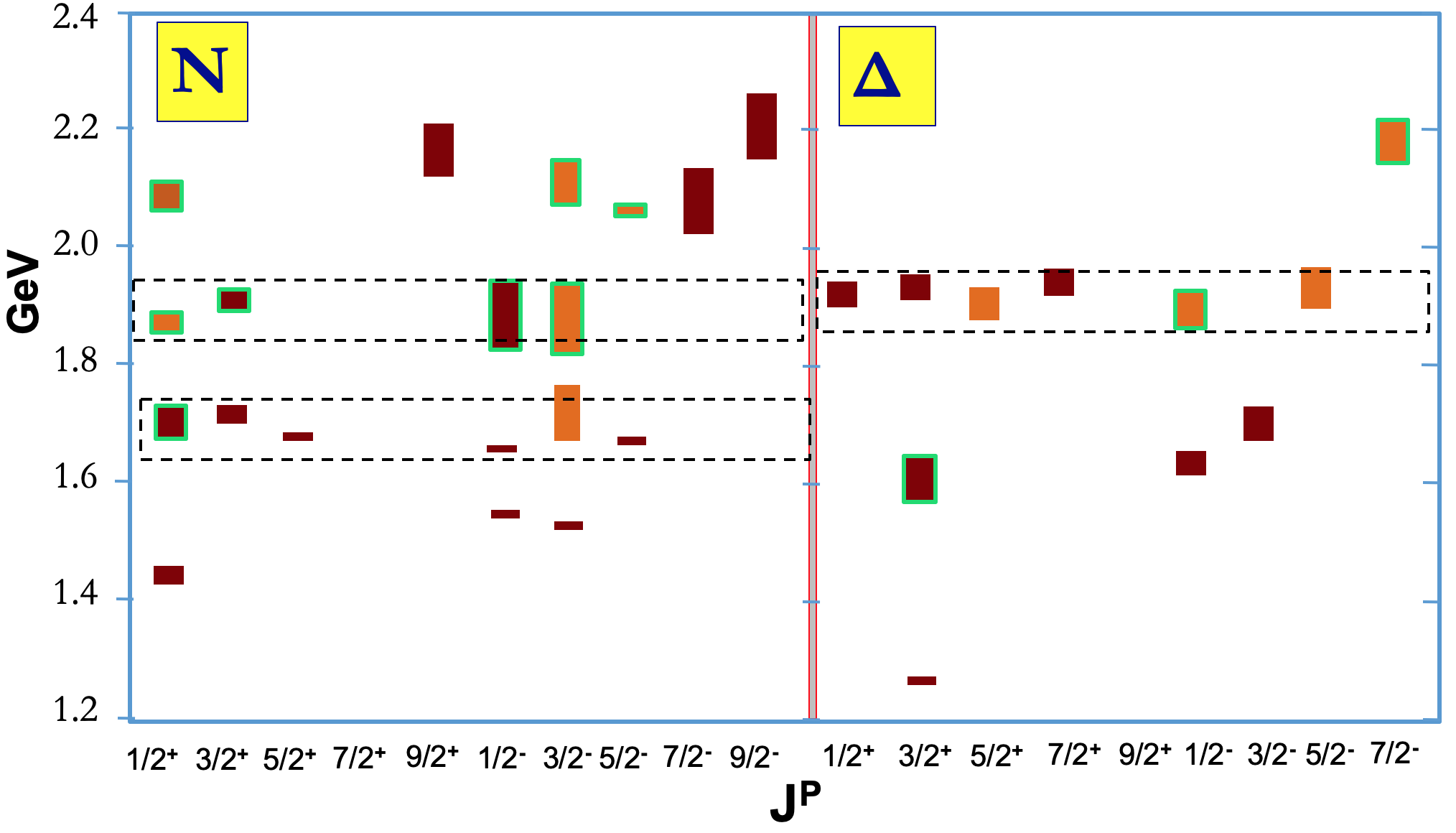}}
\caption{Top panel: Nucleon and $\Delta$ resonance spectrum below 2.2 GeV in RPP 2010~\cite{Beringer:1900zz}. 
Bottom panel: Nucleon and $\Delta$ resonance spectrum below 2.2 GeV in RPP 2018~\cite{Tanabashi:2018oca}.
The green frames highlight the new states and states with improved start ratings compared to 2010. The light brown color 
indicate 3* states, the dark color indicates 4* states. 
The dashed frames indicate apparent mass degeneracy of states with masses near 1.7 GeV and 1.9 GeV and 
different spin and parity.  The Bonn-Gatchina analysis includes all the $K^+\Lambda$ and $K^+\Sigma^\circ$ cross section and 
polarization data, as well as pion photoproduction data.}
\label{pdg}
\end{figure}

\begin{figure}[ht]
\resizebox{1.0\columnwidth}{!}{\includegraphics{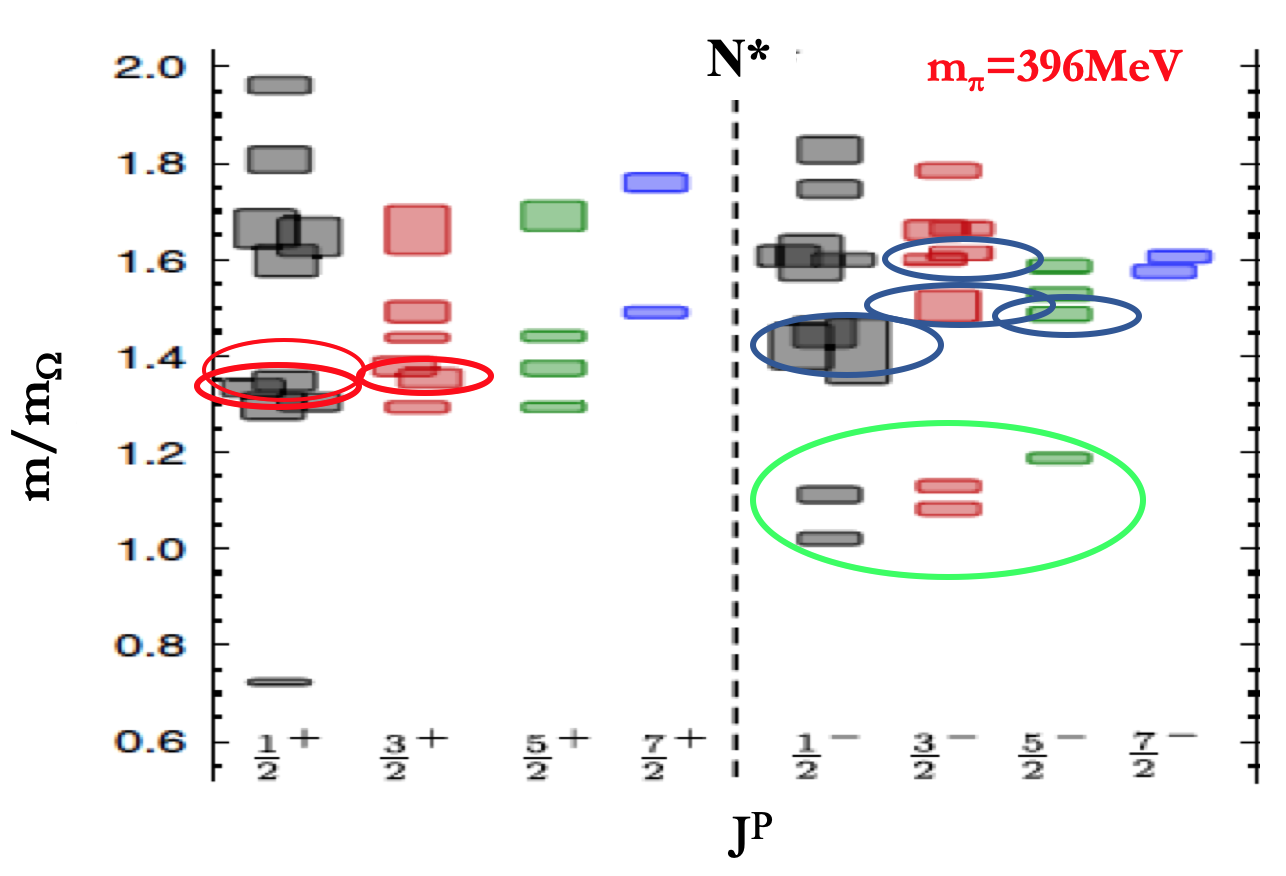}}
\caption{Nucleon resonance spectrum below 2.2 GeV from LQCD ~\cite{Edwards:2011jj}. 
The new discovered Nucleon states from RPP 2018~\cite{Tanabashi:2018oca} are indicated by the red and blue ellipses with 
their spin-parity assignments. The observed masses deviate from the predicted ones as the lattice calculation used 396MeV pion mass, 
and no multi channel coupling is included.  
The green ellipses are well-known states that come out in the LQCD work with higher masses.}
\label{lqcd}
\end{figure}

%\end{document} 

\subsection{New states from open strangeness photoproduction}
 \label{KLambda}
\begin{figure*}[ht]
\resizebox{2.0\columnwidth}{!}{\includegraphics{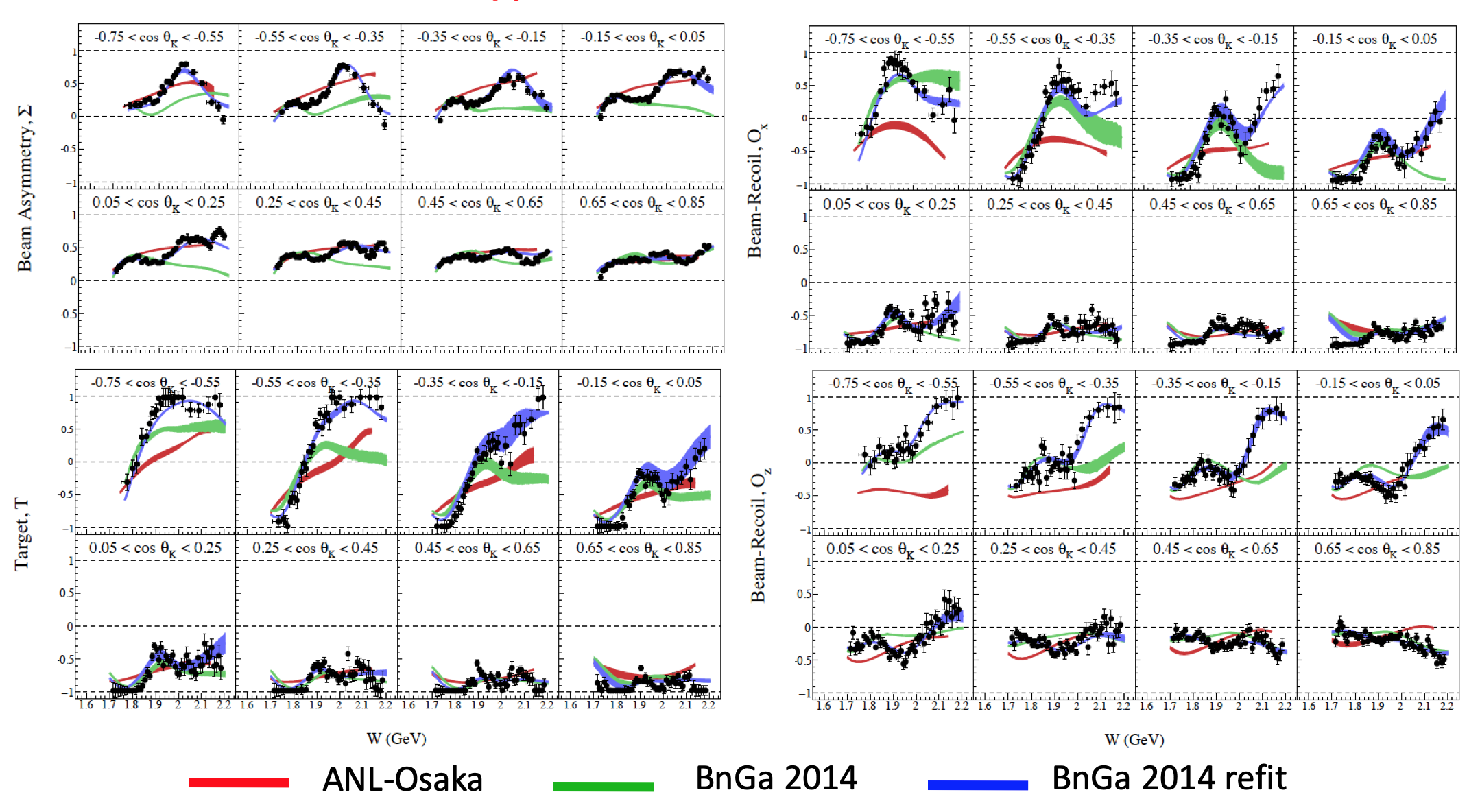}}
\caption{Beam and Target polarization asymmetries (left panel), and beam-recoil polarization asymmetries (right panel) measured 
by the CLAS collaboration in the $\gamma p \to K^+\Lambda$ channel. The projections from earlier analyses. shown in the 
red and green bands, show large discrepancies with the data in the higher W range. A refit by the BnGa group (blue band) shows that 
the discrepancies do not require new excited states but could be accommodated by just adjusting some parameters.  }
\label{Lambda-asymmetries}
\end{figure*}

 Here one focus has been on measurements of $\gamma p \to K^+\Lambda$ and $\gamma p \to K^+\Sigma$. Using 
 a linearly polarized photon beam several polarization observables can be measured by analyzing the 
 parity violating decay of the recoil $\Lambda \to p \pi^-$. It is well known that the energy-dependence of 
 a partial-wave amplitude for one particular channel is influenced by other reaction 
channels due to unitarity constraints. To fully describe the energy-dependence 
of a production amplitude other reaction channels must be included in a coupled-channel approach. 
Such analyses have been developed by the Bonn-Gatchina group\cite{Anisovich:2011fc}, 
at JLab\cite{JuliaDiaz:2007kz}, at J\"ulich\cite{Ronchen:2014cna} and other groups.   More recent 
measurements of single and double polarization data in the same channel~\cite{Paterson:2016vmc} are 
shown in Fig.~\ref{Lambda-asymmetries}.

The data sets with the highest impact on resonance amplitudes in the mass range above 1.7~GeV have been 
kaon-hyperon production using a spin-polarized photon beam and the polarization of the $\Lambda$ or $\Sigma^\circ$ 
is also measured by analyzing the parity violating decay $\Lambda \to p \pi^-$. The high precision cross section and 
polarization data~\cite{Bradford:2006ba,Bradford:2005pt,McCracken:2009ra,Dey:2010hh,McNabb:2003nf} provide 
nearly full polar angle coverage and span the $K^+\Lambda$ invariant mass range 
from threshold to 2.9 GeV, hence covering the full nucleon resonance domain where new states could be discovered. 

The backward angle $K^+\Lambda$ data in Fig.\ref{KLambda-crs} show clear resonance-like structures 
at 1.7 GeV and 1.9 GeV that are particularly prominent and well-separated from other structures at backward angles, while 
at more forward angles (not shown) t-channel processes become prominent and dominate the cross section.
The broad enhancement at 2.2~GeV may also indicate resonant behavior although it is less visible at more 
central angles with larger background contributions. 
The $K^+\Sigma$ channel also indicates significant resonant behavior as seen in Fig.~\ref{KSigma-crs}. 
The peak structure at 1.9 GeV is 
present at all angles with a maximum strength near 90 degrees, consistent with the behavior of a $J^P= {3\over 2}^+$ 
p-wave. Other structures near 2.2 to 2.3~GeV are also visible. 
Still, only a full partial wave analysis can determine the underlying resonances, their masses and spin-parity.  
The task is somewhat easier for the $K\Lambda$ channel, as the iso-scalar nature of the $\Lambda$ selects 
isospin-${1\over 2}$ states to contribute to the $K\Lambda$ 
final state, while both isospin-${1\over 2}$ and isospin-${3\over 2}$ states can contribute to the $K\Sigma$ final state.

These cross section data together with the $\Lambda$ and $\Sigma$ recoil polarization and polarization transfer data 
to the $\Lambda$ and $\Sigma$ had strong impact on the discovery of several new nucleon 
states~\cite{Anisovich:2017bsk}. They also provided new evidence for several candidate states that had been observed previously but lacked 
confirmation as shown in Fig.~\ref{pdg2014}. It is interesting to observe that four of the observed nucleon states have nearly 
degenerate masses near 1.9~GeV, as seen in Fig.~\ref{pdg}. 
Similarly, the new $\Delta$ state appears to complete a mass degenerate multiplet near 1.9~GeV as well. There is no obvious 
mechanism for this apparent degeneracy.  Nonetheless, all new states may be accommodated within the symmetric 
constituent quark model based on $SU(6)\otimes O(3)$ symmetry group as far as quantum numbers are concerned. As discussed 
in section~\ref{intro} for the case of the Roper resonance $N(1440){1\over 2}^+$, the masses of all pure  quark model states 
need to be corrected for dynamical coupled channel effects to compare them with observed resonances.  The same applies to 
the Lattice QCD predictions~\cite{Edwards:2011jj} for the nucleon and Delta spectrum.

\begin{figure}[t!]
\centering
\resizebox{1.0\columnwidth}{!}{\includegraphics{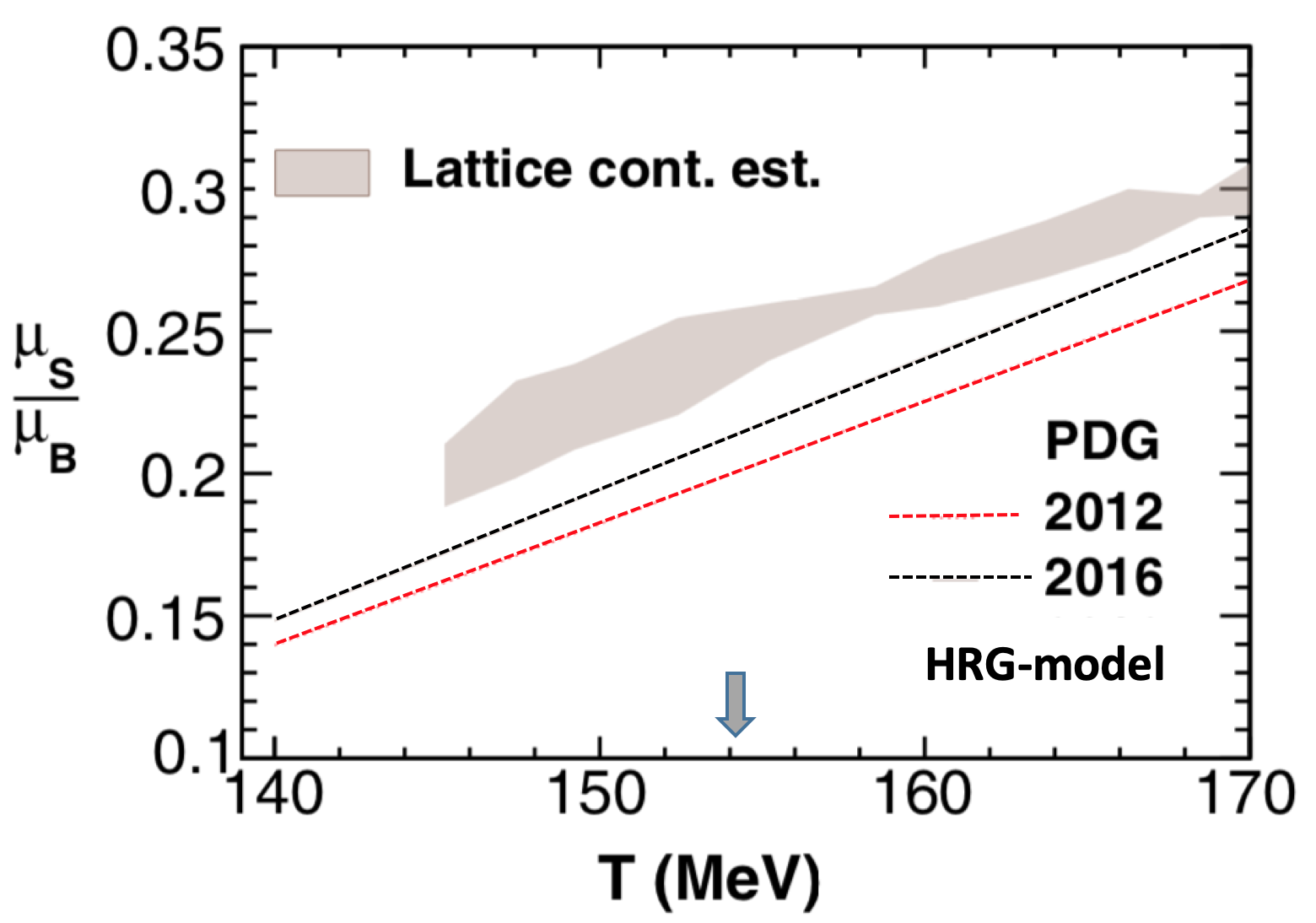}}
\caption{The ratio of baryon chemical potential of strangeness versus all baryons for the RPP 2014 and RPP 2016. The hashed 
grey area show the LQCD calculation in "hot QCD". The straight lines are calculations within a hadron gas model. The 2016 line,  
which includes more N/$\Delta$ baryon states, moves closer to the LQCD area. Note that only 3* and 4* states are included. 
If the newly discovered states in RPP 2018 (seven states that are now at 3* or 4* status) were included this line would be moving even closer to the LQCD area. }
\label{chemical-potential}
\end{figure}

Coming back to the evolution of the early universe, we may check what is the expected effect of these newly discovered 
states on the evolution of the universe near the cross over temperature 
$T_c \approx 155$~MeV? Figure~\ref{chemical-potential} shows the ratio of the baryon chemical potential of strangeness 
carrying baryons over all baryons versus temperature near the crossover temperature between de-confined and confined 
and conditions. The graph clearly shows the significant impact of the recently included new baryons by the PDG in the 2016 
edition of  the RPP in comparison to the 2012 edition. Furthermore, adding new states found since 2016 and now included in 
the 2018 edition of RPP will bring the HRG model closer to the Lattice QCD calculation~\cite{Chatterjee:2017yhp}, demonstrating 
the strong impact 
excited nucleon states have in the transition from the phase of free colored quarks and gluons to quarks and gluons confined in 
protons and neutrons.~\footnote{Increasing the number of excited baryons will lower ${\rm \mu_B}$ at a given temperature, and hence increase the ratio of $\rm{\mu_S \over \mu_B}$}     

\begin{figure}[ht]
\centering
\resizebox{1.0\columnwidth}{!}{\includegraphics{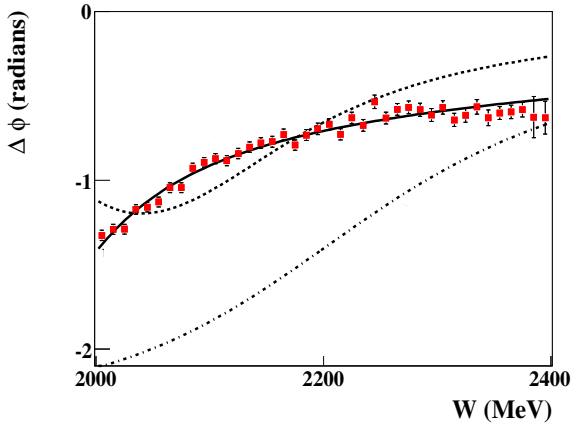}}
\caption{Phase motion of the partial wave fit to the $\gamma p \to p \omega$ differential cross section and spin density matrix elements. 3 resonant 
states, the subthreshold resonance $N(1680){5\over 2}^+$, $N(2190){7\over 2}^-$, and the missing $N(2000){5\over 2}^+$ are needed to fit the data (solid line). Fits without $N(2000){5\over 2}^+$ (dashed-dotted line), or without $N(1680){5\over 2}^+$ (dashed line) cannot reproduce the data.}
\label{omega}
\end{figure}
\begin{figure}[t!]
\resizebox{1.0\columnwidth}{!}{\includegraphics{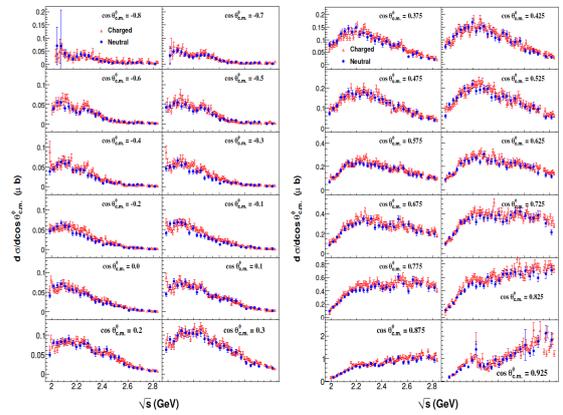}}
\caption{Differential cross sections in a nearly full angular range for $\gamma p \to p \phi$  production. }
\label{phi}
\end{figure}

\begin{figure}[t!]
\vspace{-0.8cm}\hspace{-1.2cm}\resizebox{1.3\columnwidth}{!}{\includegraphics{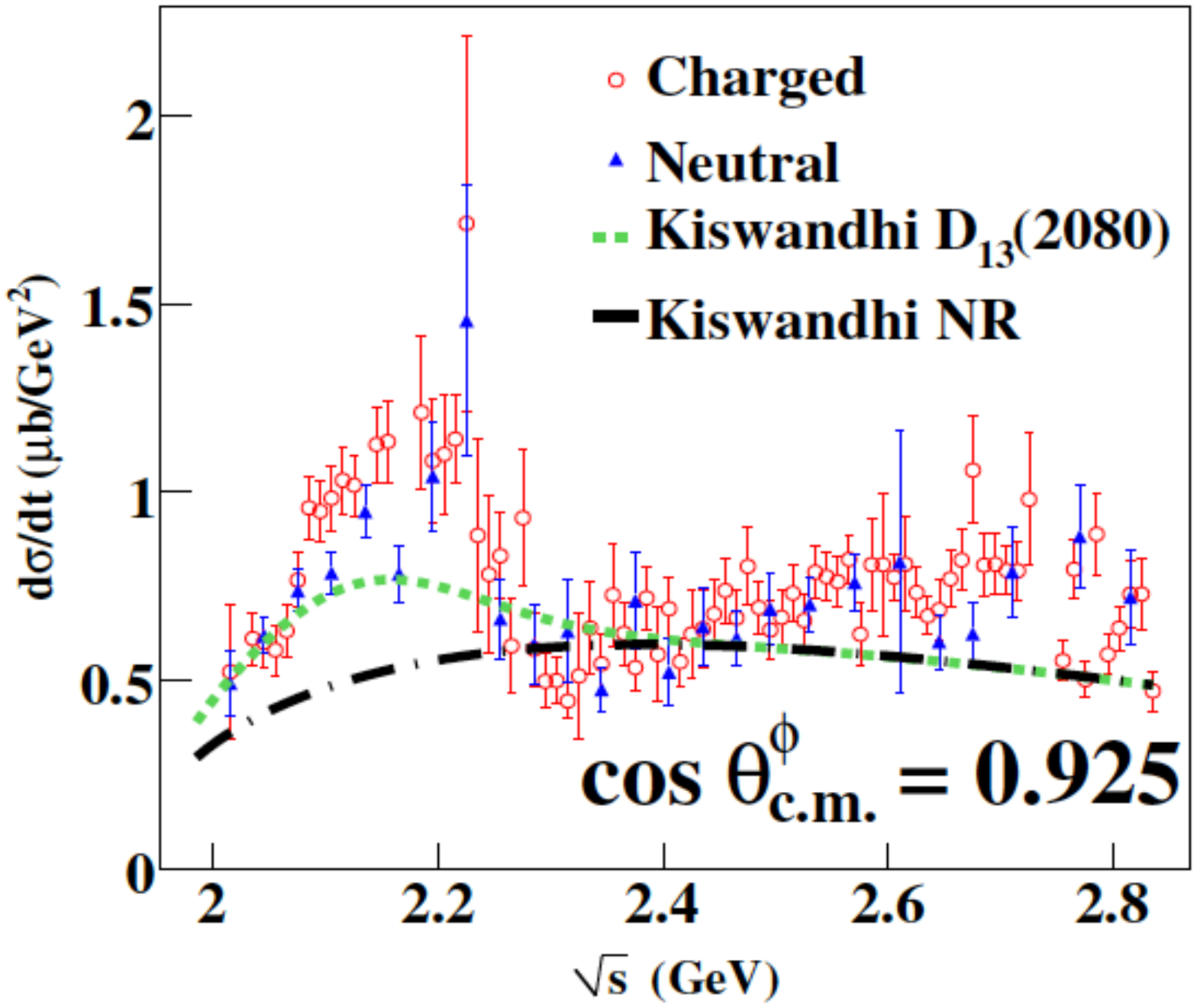}}
\caption{Differential cross sections of $\gamma p \to p \phi$  production for the most forward angle bin. The two curves 
refer to fits without (dashed) and with (dotted) a known resonance at 2.08 GeV included. }
\label{phi_forward}
\end{figure}
%\vspace{-1cm}

\subsection{Vectormeson photoproduction}
\label{Vectormeson}
Double-polarization measurements of single pion photoproduction data~\cite{Strauch:2015zob} have contributed significantly to the discovery of the high-mass $\Delta(2200)7/2^-$ state~\cite{Anisovich:2015gia}, although the state couples just at the 3.5\% level to 
N$\pi$ 
Nevertheless in the mass range above 2.0~GeV resonances tend to decouple from simple final states like $N\pi$, $N\eta$, and 
$K\Lambda$, the search for undiscovered high-mass states requires to consider more complex final states 
with multi-mesons $N\pi\pi$ or vector mesons, such 
as $N\omega$, $N\phi$, and $K^*\Sigma$. The study of such final states adds significant complexity as many more amplitudes can 
contribute to these photoproduction processes compared to single pseudo-scalar meson production. As is the case for $N\eta$ production, 
the $N\omega$ channel is selective to isospin $1\over 2$ nucleon states only.       
The CLAS collaboration has collected a tremendous amount of data in the $p\omega$ channel~\cite{Williams:2009aa,Williams:2009ab}, including 
single and double polarization measurements~\cite{Collins:2017vev,Akbar:2017uuk,Roy:2017qwv,Roy:2018tjs,Anisovich:2017rpe}.  The CLAS collaboration performed a single channel 
event-based analysis, whose results are shown in Fig.~\ref{omega}, and provided further evidence for the $N(2000){5\over 2}^+$. Also 
a large amount of $p\phi$~\cite{Seraydaryan:2013ija,Dey:2014tfa} final states on differential cross sections and spin-density matrix elements have been published, although they have not systematically included in the more complex multi-channel analyses.  
Photoproduction of $\phi$ mesons is also considered a potential source of new excited nucleon states in the mass range above 2 GeV. 
Differential cross sections and spin-density matrix elements 
have been measured for $\gamma p \to p \phi$ in a mass range up to nearly 3 GeV. In Fig.~\ref{phi} structures are seen near 2.2~GeV in the forward most angle bins and at very backward angles for both decay channels $\phi \to K^+K^-$ and 
$\phi \to K_l^0K_s^0$, and with the exception of the smallest forward angle bin the structures are more prominent at backward angles. 
Only a multi-channel partial wave analysis will be able to pull out any significant resonance strength.  
Fig.~\ref{phi_forward} shows the differential cross section 
$d\sigma /dt$ 
of the most forward angle bin. A broad structure at 2.2 GeV is present, but does not show the typical Breit-Wigner behavior of a single 
resonance. It also does 
not fit the data in a larger angle range, which indicates that contributions other than genuine resonances may be significant. 
The forward and backward angle structures 
may also hint at the presence of dynamical effects possibly due to molecular contributions such as diquark-anti-triquark 
contributions~\cite{Lebed:2015dca}, the strangeness equivalent to the recently observed hidden charm $P_c^+$ states.   
 \begin{figure}[t!]
\centerline{\resizebox{1.0\columnwidth}{!}{\includegraphics{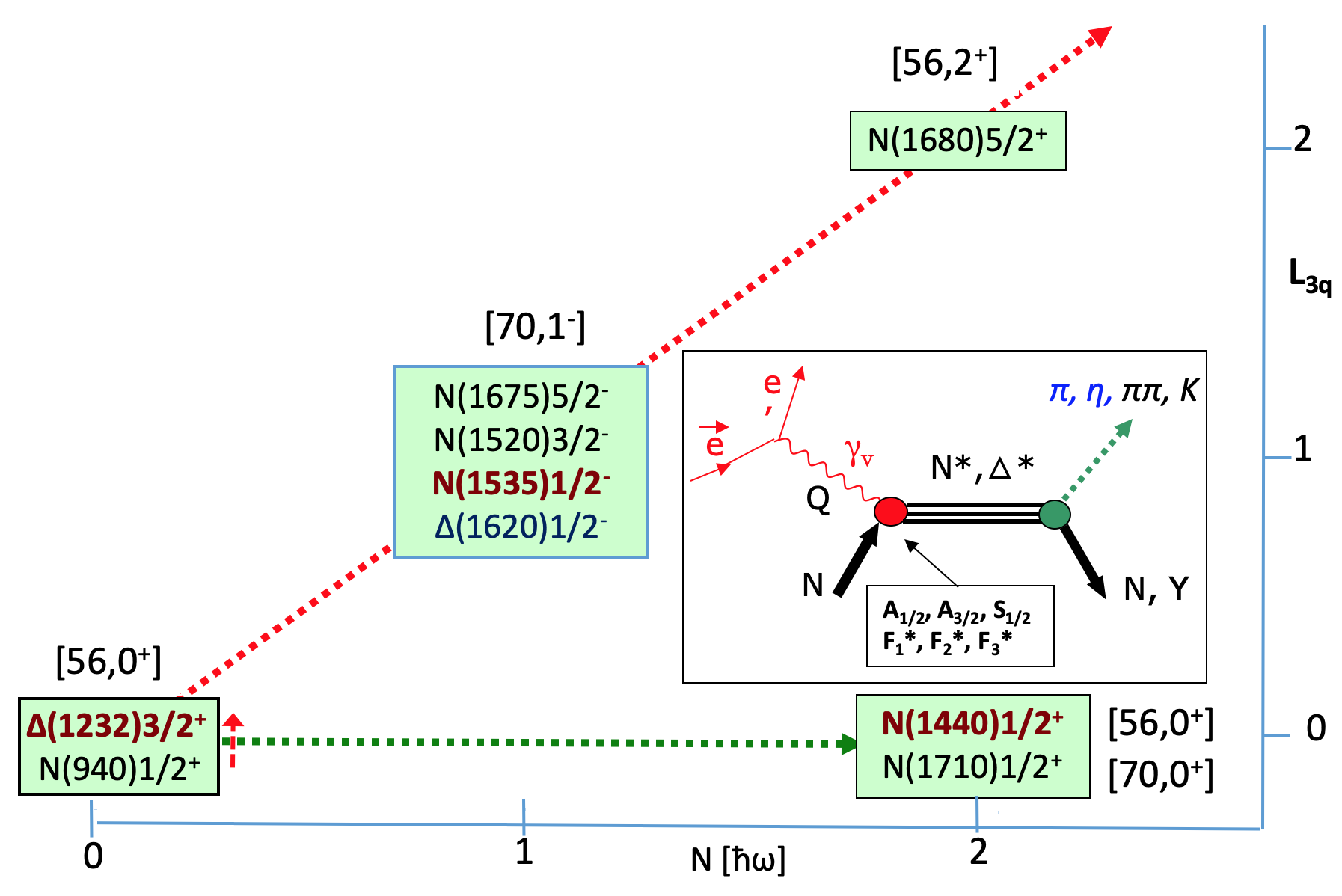}}}
\caption{Schematics of $SU(6)\otimes O(3)$ supermultiplets with some of the excited states that have been explored in 
$ep\to e^\prime \pi^+ n$ ,$ep\to e^\prime p \pi^\circ$, $ep\to e^\prime p \eta$, and $ep \to e^\prime p \pi^+ \pi^-$ 
electroproduction experiments. The inset shows the helicity amplitudes and electromagnetic multipoles often used to 
describe the data. Only the states highlighted in red are discussed here.}
\label{SU6}
\end{figure}   
Another process that has promise in the search for new excited baryon states, including those with 
isospin-${3\over 2}$ is $\gamma p \to K^*\Sigma$~\cite{sarantsev2015}. 
In distinction to the vector mesons discussed above, diffractive processes do not play a role in this channel, which then 
may allow better direct access to s-channel resonance production.

\begin{figure}[t!]
\centerline{\resizebox{1.0\columnwidth}{!}{\includegraphics{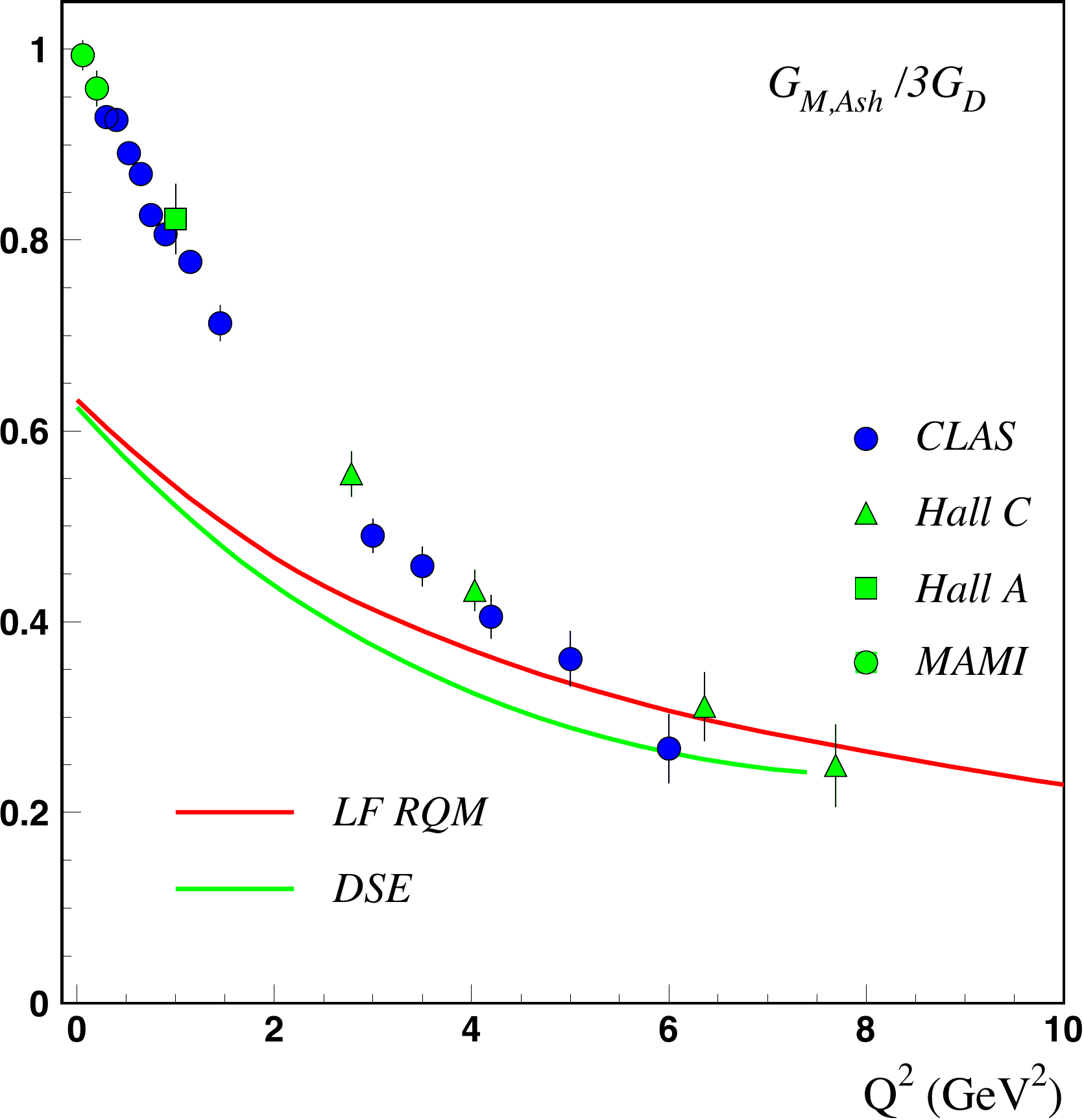}}}
\caption{(Color online) The magnetic N$\Delta$ transition form factor normalized to the dipole form.factor compared with the LF RQM with running
quark mass, and with DSE/QCD. Both are close to the data at high $Q^2$. At $Q^2 < 3$GeV$^2$ meson-baryon contributions are significant.  }
\label{Delta-Gmn}
\vspace{0.5cm} 
%\end{figure}
%\begin{figure}[th!]
\centerline{\resizebox{1.0\columnwidth}{!}{\includegraphics{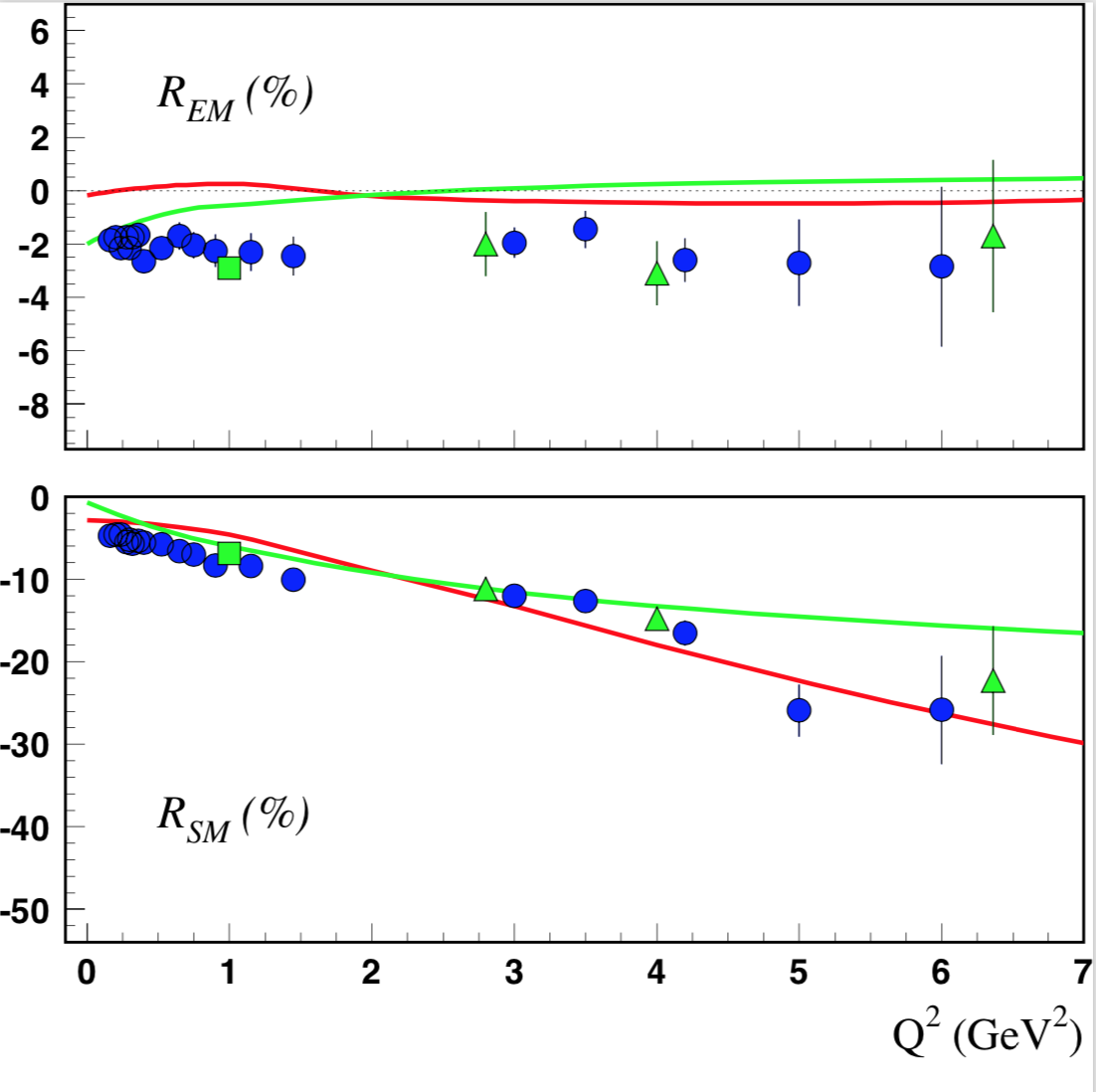}}}
\caption{(Color online) The ratios $R_{EM}$ and $R_{SM}$ for the $N\Delta(1232)$ transition.  
The solid red curves are the LF RQM predictions, the green curves are from the DSE approach. References to data as in Fig.~\ref{Delta-Gmn}.}
\label{Delta-ratios}
\end{figure}

\section{Structure of excited nucleons}
\label{structure}
\begin{figure}[t!]
\centering
\resizebox{1.0\columnwidth}{!}{\includegraphics{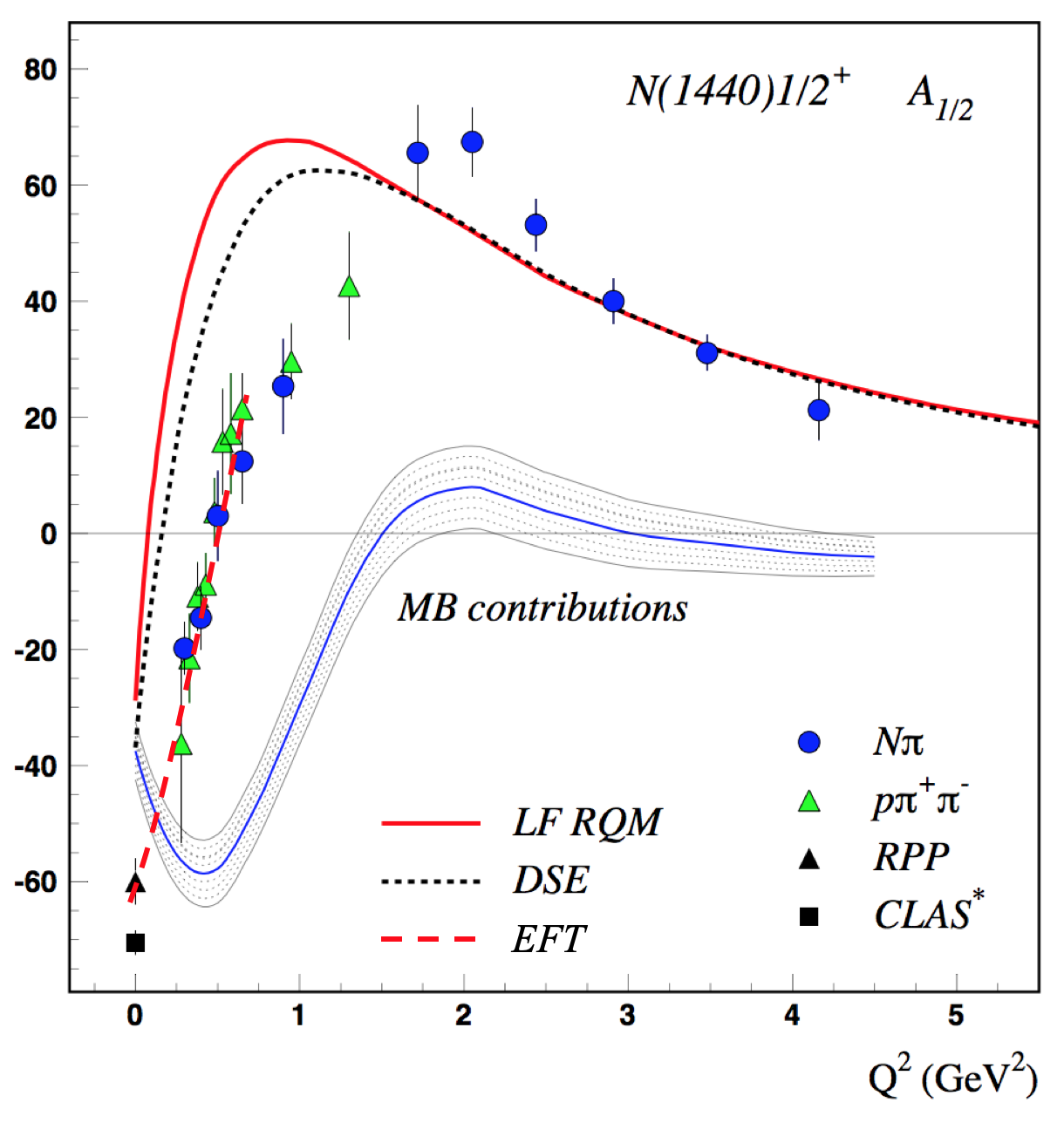}}
\caption{The transverse helicity amplitudes $A_{1/2}$ for the Roper resonance $N(1440){1\over 2}^+$. Data are from CLAS compared
to LF RQM with running quark masses (solid line), and with projections from the DSE/QCD approach (dotted line). The 
shaded band indicates non 3-quark contributions inferred from the difference of the LF RQM curve and the CLAS data. 
The red dashed line is the EFT calculation that describes the data at small $Q^2$.}
\label{p11}
%\end{figure}
%\begin{figure}[ht!]
\vspace{0.5cm}
\centering
\resizebox{1.0\columnwidth}{!}{\includegraphics{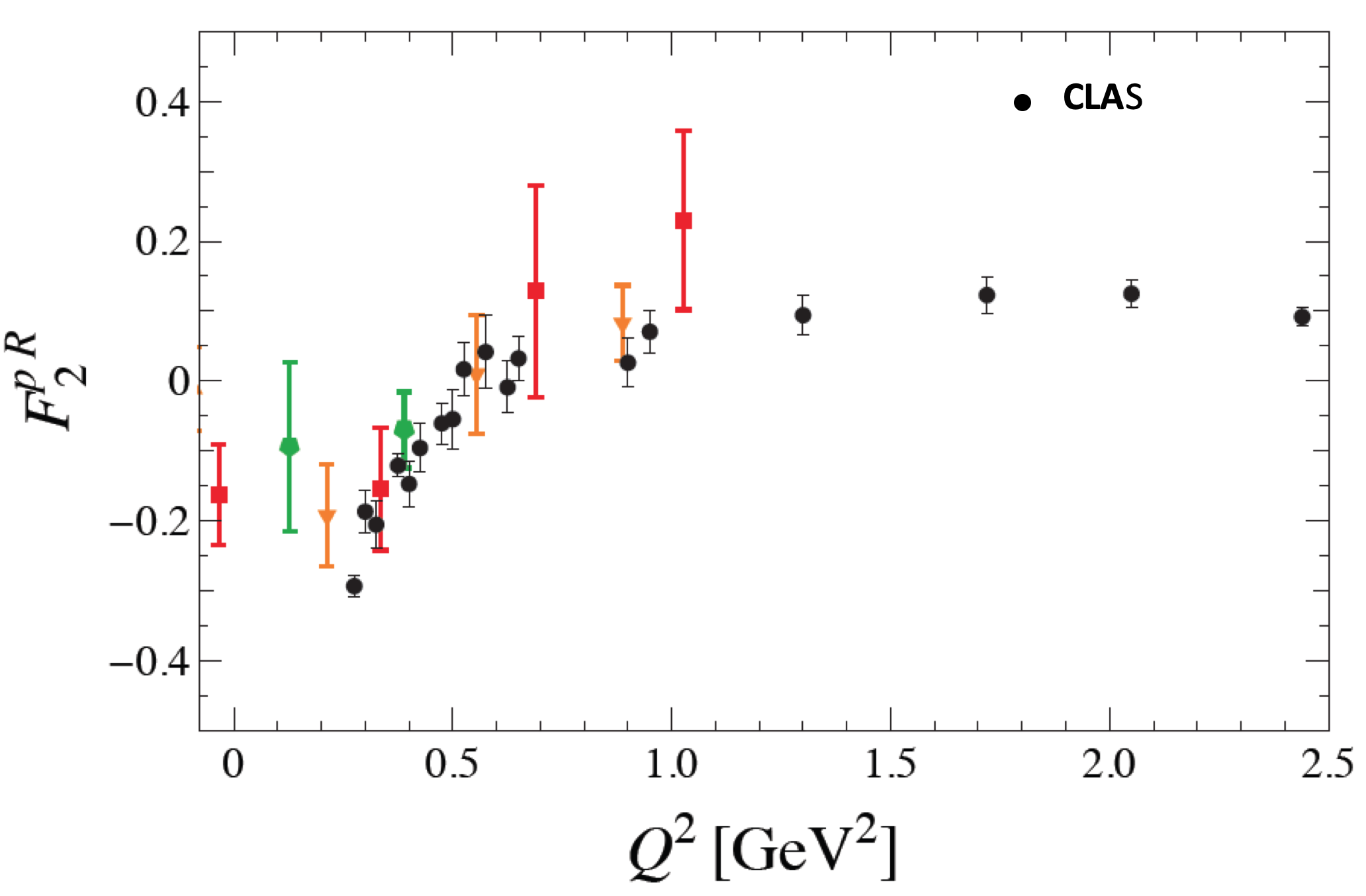}}
\caption{The $pN^+(1440)1/2^+$ transition amplitude $F_2(Q^2)$ from LQCD~\cite{Lin:2011da} compared to CLAS results. }
\label{Roper-LQCD}
\end{figure}

\begin{figure}[thb!]
\centering
\resizebox{1.0\columnwidth}{!}{\includegraphics{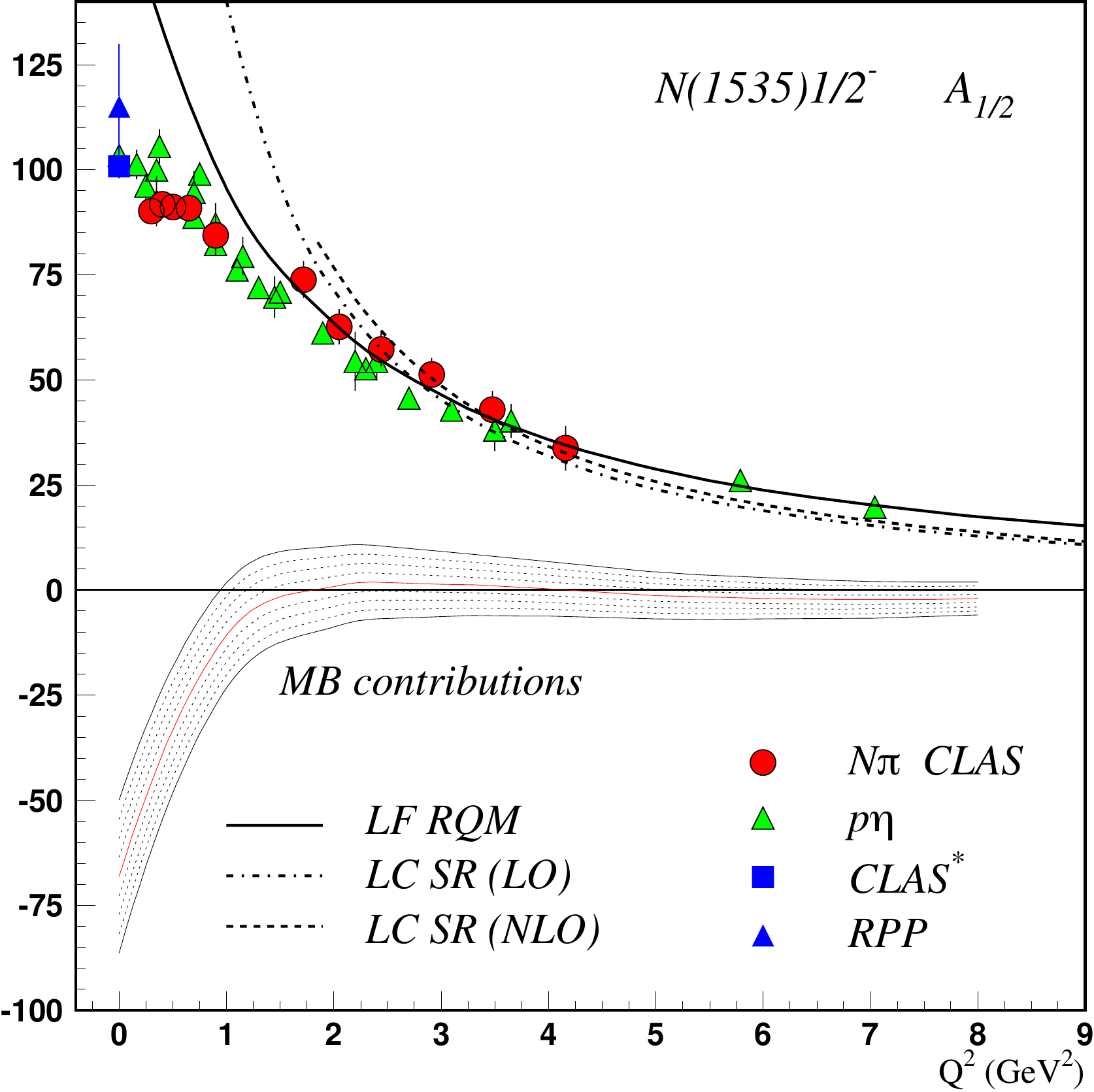}}
\caption{The transverse amplitude $A_{1/2}$ for the $N(1535){1\over 2}^-$ resonance compared to LF RQM calculations (solid line) 
and QCD computation within the LC Sum Rule approach.}
\label{s11}
\end{figure}

Meson photoproduction has become an essential tool in the search for new excited baryons. The exploration of the internal structure 
of excited states and the effective degrees of freedom contributing to s-channel resonance excitation requires the use of electron beams, 
where the virtuality ($Q^2$) of the exchanged photon can be varied to probe the spatial structure. 
Electroproduction of final states with pseudoscalar mesons 
(e.g. $N\pi$, $p\eta$, $K\Lambda$) have been employed with CLAS, leading to new insights into the scale dependence of 
effective degrees of freedom, e.g. meson-baryon, constituent quark, and dressed quark contributions. Several excited states, shown 
in Fig.~\ref{SU6} assigned to their primary $SU(6) \otimes O(3)$ supermultiplets have been studied. 
The $N\Delta(1232){3\over 2}^+$ 
transition is now well measured in a large range of $Q^2$~\cite{Joo:2001tw,Ungaro:2006df,Frolov:1998pw}. Results on the 
magnetic transition form factor $\rm G_{Mn}$ and on the quadrupole transition ratios are shown in Fig.~\ref{Delta-Gmn} and 
Fig.~\ref{Delta-ratios}.  Two of the prominent higher mass states, the Roper resonance
$N(1440){1\over 2}^+$  and $N(1535){1\over 2}^-$ are shown in Fig.~\ref{p11} and in Fig.~\ref{s11}, respectively, as representative 
examples~\cite{Aznauryan:2008pe,Aznauryan:2009mx,Burkert:2019bhp} of a wide program at 
JLab~\cite{Mokeev:2012vsa,Mokeev:2015lda,Fedotov:2018oan,Isupov:2017lnd,Denizli:2007tq,Armstrong:1998wg,Egiyan:2006ks,Park:2007tn,Park:2014yea}. 
For these three states advanced relativistic quark model calculations~\cite{Aznauryan:2015zta} and QCD calculations from DSE~\cite{Segovia:2015hra} and 
Light Cone sum rule~\cite{Anikin:2015ita} are available, for the first time employing QCD-based modeling of the excitation 
of the quark core. 

There is agreement with the data at $Q^2 > 1.5$~GeV$^2$ for the latter two states, while the meson-baryon contributions for 
the $\Delta(1232)$ are more extended, and agreement with the quark based calculations is reached at $Q^2 > 4$~GeV$^2$.  The calculations deviate significantly from the data 
at lower $Q^2$, which indicates significant  non quark core effects.  For the Roper resonance such contributions have been described 
successfully in dynamical meson-baryon models~\cite{Obukhovsky:2011sc} and in effective field theory~\cite{Bauer:2014cqa}. Calculations on the Lattice for the N-Roper transition amplitudes have been carried out with dynamical quarks~\cite{Lin:2011da}. The results agree 
with the data in the range $Q^2 < 1.0$~GeV$^2$, where data and calculations overlap~Fig.~\ref{Roper-LQCD}.

\begin{figure}[t!]
%\vspace{-0.5cm}
\centerline{\resizebox{0.9\columnwidth}{!}{\includegraphics{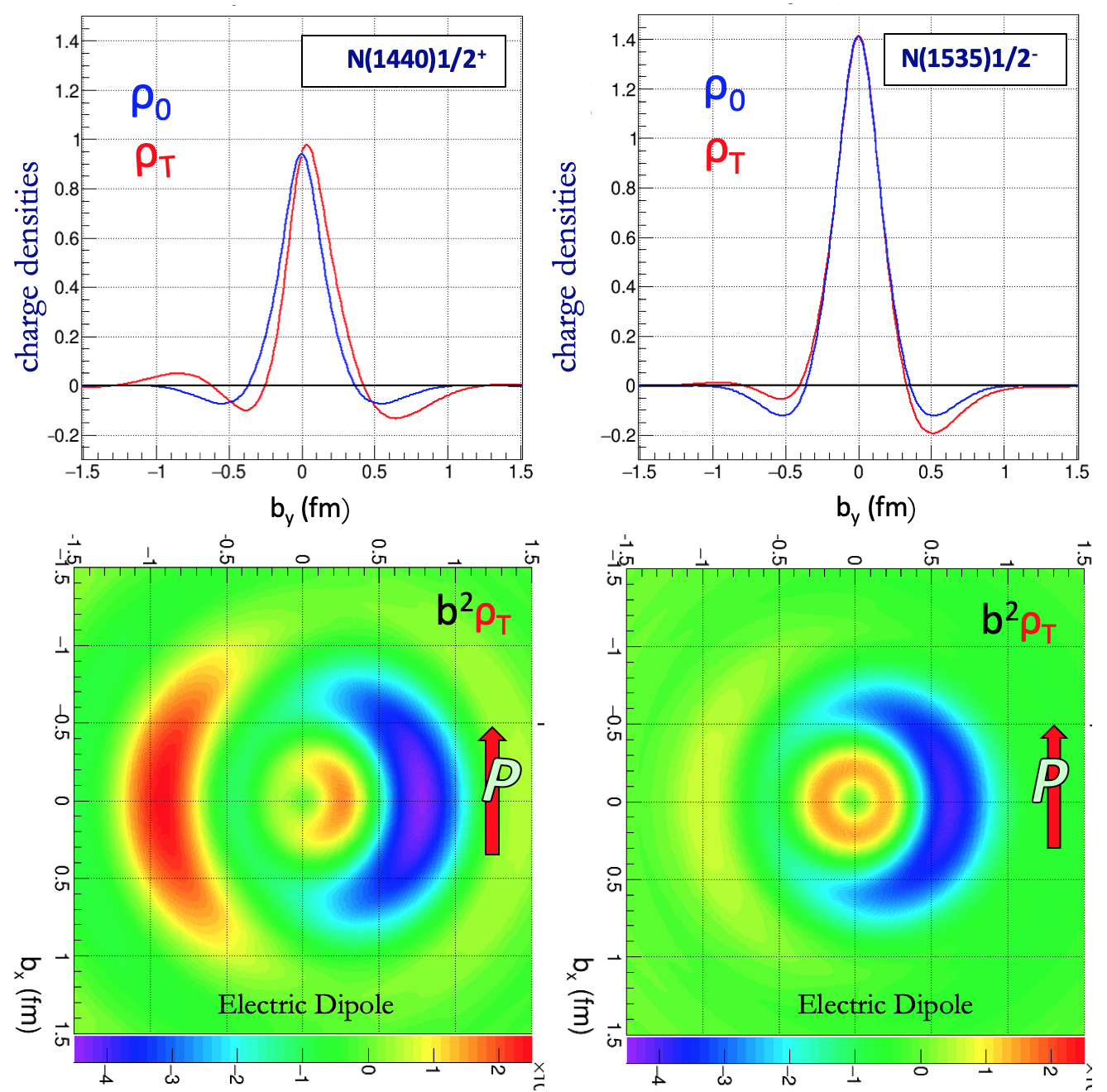}}}
\caption{Left panels: $N(1440)$, top: projection of charge densities on $b_y$, bottom: transition charge densities when the proton  
is spin polarized along $b_x$. Right panels: same for $N(1535)$. Note that the densities are scaled with 
$b^2$ to emphasize the outer wings. Color code:negative charge is blue, positive charge is red. Note that all scales
are the same. }
\label{charge_densities}
\end{figure}

Knowledge of the helicity amplitudes in a large $Q^2$ allows for the determination of the transition charge densities on the light 
cone in transverse impact parameter space ($b_x, b_y$)~\cite{Tiator:2008kd}. Figure~\ref{charge_densities} shows the comparison 
of $N(1440){1\over 2}^+$ and  $N(1535){1\over 2}^-$. There are clear differences in the charge transition densities between 
the two states. The Roper state has a softer positive core and a wider negative outer cloud than $N(1535)$ 
and develops a larger shift in $b_y$ when the proton is polarized along the $b_x$ axis.     
\begin{figure}[t!]
\centering
%\vspace{1cm}
\resizebox{0.48\columnwidth}{!}{\includegraphics{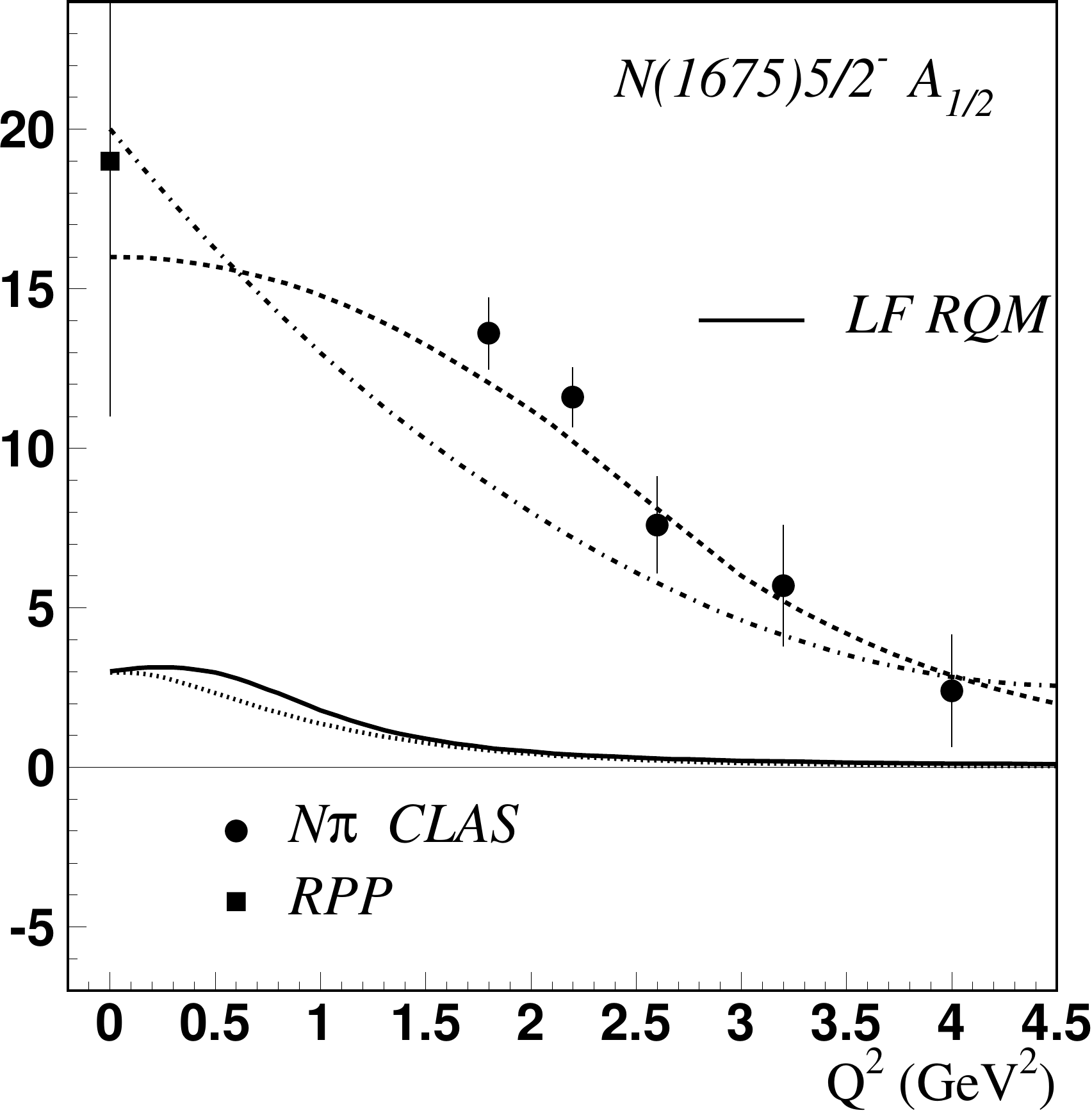}}
 \resizebox{0.48\columnwidth}{!}{\includegraphics{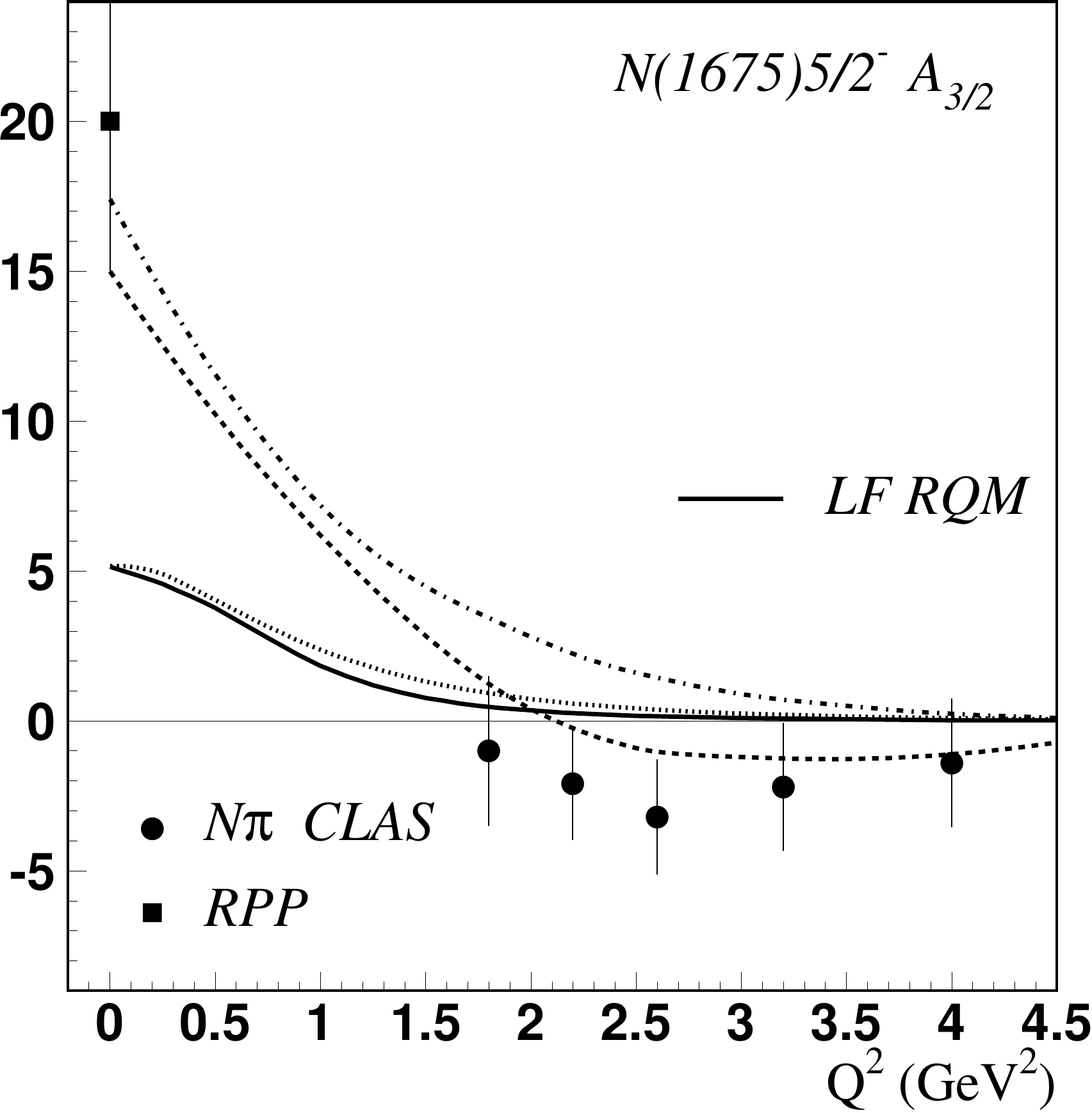}}
\caption{Helicity amplitude $A_{1/2}$ (left) and $A_{3/2}$ (right) for $N^+(1675){5/2}^-$ off proton target. }
\label{N1675}
\end{figure}
New electroproduction data on the Roper and several higher mass states have been obtained in the 2-pion channel, 
specifically in $ep\to e'p\pi^+\pi^-$~\cite{Mokeev:2015lda}. 

\subsection{The $N(1675){5/2}^-$  state - revealing the meson-baryon contributions}
In previous discussions we have assumed that meson-baryon degrees of freedom provide significant strength to the 
resonance excitation in the low $Q^2$ domain where quark based approaches LF RQM, DSE/QCD, and LCSR calculations fail to 
reproduce the transition amplitudes quantitatively. Our conclusion rests, in part, with this assumption. 
But, how can we be certain of the validity of this assumption? 

The $N(1675){5/2}^-$ resonance allows testing this assumption, quantitatively. Figure~\ref{N1675} shows 
our current knowledge of the transverse helicity amplitudes $A_{1/2}(Q^2)$ and $A_{1/2}(Q^2)$, for proton target compared to  
RQM~\cite{Aznauryan:2017nkz} and 
hypercentral CQM~\cite{Santopinto:2012nq} calculations. The specific quark transition for a $J^P = 5/2^-$ state belonging to the 
$SU(6)\otimes O(3)] = [70, 1^-]$ supermultiplet configuration, in non-relativistic approximation prohibits the transition from the
 proton in a single quark transition.
This suppression, known as the Moorhouse selection rule~\cite{Moorhouse:1966jn}, is valid for the transverse transition 
amplitudes $A_{1/2}$ and $A_{3/2}$ at all $Q^2$. It should be noted that this selection rule does apply to the 
transition from protons but not from neutrons. Modern quark models, that go beyond single quark transitions,  
confirm quantitatively the suppression resulting in very 
small transition amplitudes from protons but large ones from neutrons. The measured helicity amplitudes off the protons are almost  
exclusively due to meson-baryon contributions as the dynamical coupled channel (DCC) calculation indicates (dashed line). 
The close correlation of the 
DCC calculation and the measured data for the case when quark contributions are nearly absent,    
supports the phenomenological description of the helicity amplitudes in terms of a 3-quark core that dominate at high $Q^2$ and 
meson-baryon contributions that can make important contributions at lower $Q^2$.

\section{Conclusions and Outlook}
Over the past five years eight baryon states in the mass range from 1.85 to 2.15 GeV have been either discovered or 
evidence for the existence of states has been significantly strengthened. To a large degree this is the result of adding 
very precise photoproduction data in open strangeness channels to the data base that is included in multi-channel partial wave 
analyses, especially the Bonn-Gatchina PWA. The possibility to measure polarization observables in these processes has been 
critical. In the mass range above 2 GeV more complex processes such as vector mesons or $\Delta\pi$ 
may have sensitivity 
to states with higher masses but require more complex analyses techniques to be brought to bear. Precision data in such 
channels have been available for a few years but remain to be fully incorporated in multi-channel partial wave analyses processes. 
The light-quark baryon spectrum is likely also populated with hybrid excitations~\cite{Dudek:2012ag} where the gluonic 
admixtures to the wave function are dominating the excitation. These states appear with the same quantum numbers as 
ordinary quark excitations, and can only be isolated from ordinary states due to the $Q^2$ dependence of their helicity 
amplitudes~\cite{Li:1991yba}, which is expected to be quite different from ordinary quark excitations. This requires new 
electroproduction data especially at low $Q^2$~\cite{LOI_Hybrids} with different final states and at masses above 2 GeV.  
 
Despite the very significant progress made in recent years to further establish the light-quark baryon spectrum and explore 
the internal structure of excited states, much remains to be done. A vast amount of precision data already collected 
needs  to be included in the multi-channel analysis frameworks, and polarization data are still to be analyzed. 

There are approved proposals to study resonance excitations at much higher $Q^2$ and with higher precision at Jefferson Lab 
with CLAS12~\cite{elou19,Burkert:2018nvj}, which may reveal the transition to the bare quark core contributions at short distances.   

A new avenue of experimental research has recently been opened up with the data-based extraction of the first mechanical property of the proton - its internal pressure distribution~\cite{Burkert:2018bqq}. Mechanical properties of resonance transitions 
have recently been explored for the $N(1535)3/2^- \to N(938)$ gravitational transition form factors calculations~\cite{Ozdem:2019pkg}. In order to access these new gravitational form factors experimentally, the nucleon to resonance transition GPDs 
have to be studied. The framework for studying the $N \to N(1535)$ transition GPDs, which would enable experimental access to these mechanical properties, still remains to be developed.   

\vspace{0.3cm} 

\noindent I like to thank Inna Aznauryan and Viktor Mokeev for numerous discussions on the subjects discussed in this presentation.

\section{Acknowledgment}This work was supported by the US Department of Energy under contract No. DE-AC05-06OR23177.
This work was supported by
the U.S. Department of Energy, Office of Science,
Office of Nuclear Physics, under Contract
No. DE-AC05-06OR23177.


\begin{thebibliography}{999}

\footnotesize{

\bibitem{Bazavov:2014xya} 
  A.~Bazavov {\it et al.},
%  ``Additional Strange Hadrons from QCD Thermodynamics and Strangeness Freezeout in Heavy Ion Collisions,''
  Phys.\ Rev.\ Lett.\  {\bf 113}, 072001 (2014)
  doi:10.1103/PhysRevLett.113.072001
  [arXiv:1404.6511 [hep-lat]].

\bibitem{Bazavov:2014yba} 
  A.~Bazavov {\it et al.},
% ``The melting and abundance of open charm hadrons,''
  Phys.\ Lett.\ B {\bf 737}, 210 (2014)
  doi:10.1016/j.physletb.2014.08.034
  [arXiv:1404.4043 [hep-lat]].

\bibitem{Capstick:1986bm} 
  S.~Capstick and N.~Isgur,
%  ``Baryons in a Relativized Quark Model with Chromodynamics,''
  Phys.\ Rev.\ D {\bf 34}, 2809 (1986)
  [AIP Conf.\ Proc.\  {\bf 132}, 267 (1985)]. 
    doi:10.1103/PhysRevD.34.2809, 10.1063/1.35361

\bibitem{Anderson:1952nw} 
  H.~L.~Anderson, E.~Fermi, E.~A.~Long and D.~E.~Nagle (1952)
%  ``Total Cross-sections of Positive Pions in Hydrogen,''
  Phys.\ Rev.\  {\bf 85}, 936 . 
 
\bibitem{GellMann1964} 
  M. Gell-Mann, 
  Phys. Lett.  {\bf 8}, 214 (1964).

\bibitem{Zweig1964} 
  G. Zweig, CERN Reports, TH 401 and 412  (1964).

\bibitem{Greenberg:1964pe} 
  O.~W.~Greenberg,
 % ``Spin and Unitary Spin Independence in a Paraquark Model of Baryons and Mesons,''
  Phys.\ Rev.\ Lett.\  {\bf 13}, 598 (1964); arXiv:0803.0992 [physics.hist-ph].

\bibitem{Burkert:2004sk} 
  V.~D.~Burkert and T.~S.~H.~Lee 
 % ``Electromagnetic meson production in the nucleon resonance region,''
  Int.\ J.\ Mod.\ Phys.\ E {\bf 13}, 1035 (2004).

\bibitem{Beringer:1900zz} 
  J.~Beringer {\it et al.} [Particle Data Group],
 % ``Review of Particle Physics (RPP),''
  Phys.\ Rev.\ D {\bf 86}, 010001 (2012).
  doi:10.1103/PhysRevD.86.010001

\bibitem{Agashe:2014kda} 
  K.~A.~Olive {\it et al.} [Particle Data Group],
%  ``Review of Particle Physics,''
  Chin.\ Phys.\ C {\bf 38}, 090001 (2014).
  doi:10.1088/1674-1137/38/9/090001

\bibitem{Patrignani:2016xqp} 
  C.~Patrignani {\it et al.} [Particle Data Group],
% ``Review of Particle Physics,''
  Chin.\ Phys.\ C {\bf 40}, no. 10, 100001 (2016).
  doi:10.1088/1674-1137/40/10/100001

\bibitem{Tanabashi:2018oca} 
  M.~Tanabashi {\it et al.} [Particle Data Group],
 % ``Review of Particle Physics,''
  Phys.\ Rev.\ D {\bf 98}, no. 3, 030001 (2018).
  doi:10.1103/PhysRevD.98.030001


\bibitem{Dudek:2012ag} 
  J.~J.~Dudek and R.~G.~Edwards,
  Phys.\ Rev.\ D {\bf 85}, 054016 (2012)

\bibitem{Suzuki:2009nj} 
  N.~Suzuki et al., 
 % ``Disentangling the Dynamical Origin of P-11 Nucleon Resonances,''
  Phys.\ Rev.\ Lett.\  {\bf 104}, 042302 (2010)
  [arXiv:0909.1356 [nucl-th]].
  
\bibitem{Klempt:2009pi} 
  E.~Klempt and J.~M.~Richard,
 % ``Baryon spectroscopy,''
  Rev.\ Mod.\ Phys.\  {\bf 82}, 1095 (2010)

\bibitem{Tiator:2011pw} 
  L.~Tiator, D.~Drechsel, S.~S.~Kamalov and M.~Vanderhaeghen,
 % ``Electromagnetic Excitation of Nucleon Resonances,''
  Eur.\ Phys.\ J.\ ST {\bf 198}, 141 (2011).   

\bibitem{Aznauryan:2011qj} 
  I.~G.~Aznauryan and V.~D.~Burkert,
 % ``Electroexcitation of nucleon resonances,''
  Prog.\ Part.\ Nucl.\ Phys.\  {\bf 67}, 1 (2012)

\bibitem{Aznauryan:2012ba} 
  I.~G.~Aznauryan {\it et al.},
 % ``Studies of Nucleon Resonance Structure in Exclusive Meson Electroproduction,''
  Int.\ J.\ Mod.\ Phys.\ E {\bf 22}, 1330015 (2013)

\bibitem{Crede:2013sze} 
  V.~Crede and W.~Roberts,
 % ``Progress towards understanding baryon resonances,''
  Rept.\ Prog.\ Phys.\  {\bf 76}, 076301 (2013)

\bibitem{Capstick:1993kb} 
  S.~Capstick and W.~Roberts,
%  ``Quasi two-body decays of nonstrange baryons,''
  Phys.\ Rev.\ D {\bf 49}, 4570 (1994)
  
\bibitem{Dugger:2005my} 
  M.~Dugger {\it et al.},
 % ``Eta-prime photoproduction on the proton for photon energies from 1.527-GeV to 2.227-GeV,''
  Phys.\ Rev.\ Lett.\  {\bf 96}, 062001 (2006)
  [Phys.\ Rev.\ Lett.\  {\bf 96}, 169905 (2006)]

\bibitem{Dugger:2009pn} 
  M.~Dugger {\it et al.} [CLAS],
 % ``pi+ photoproduction on the proton for photon energies from 0.725 to 2.875-GeV,''
  Phys.\ Rev.\ C {\bf 79}, 065206 (2009)
  doi:10.1103/PhysRevC.79.065206

\bibitem{Mattione:2017fxc} 
  P.~T.~Mattione {\it et al.} [CLAS],
%  ``Differential cross section measurements for $\gamma n\rightarrow{\pi}^{-}p$ above the first nucleon resonance region,''
  Phys.\ Rev.\ C {\bf 96}, no. 3, 035204 (2017)
  doi:10.1103/PhysRevC.96.035204

\bibitem{Senderovich:2015lek}
  I.~Senderovich {\it et al.} [CLAS],
 % ``First measurement of the helicity asymmetry $E$ in $\eta$ photoproduction on the proton,''
  Phys.\ Lett.\ B {\bf 755} (2016) 64
  doi:10.1016/j.physletb.2016.01.044

\bibitem{Williams:2009yj} 
  M.~Williams {\it et al.} [CLAS Collaboration],
 % ``Differential cross s ections for the reactions gamma p ---> p eta and gamma p ---> p eta-prime,''
  Phys.\ Rev.\ C {\bf 80}, 045213 (2009)

\bibitem{Collins:2017sgu} 
  P.~Collins {\it et al.} [CLAS],
 % ``Photon beam asymmetry $\Sigma$ for $\eta$ and $\eta^\prime$ photoproduction from the proton,''
  Phys.\ Lett.\ B {\bf 771}, 213 (2017)
  doi:10.1016/j.physletb.2017.05.045

 \bibitem{Williams:2009aa} 
  M.~Williams {\it et al.} [CLAS],
 % ``Partial wave analysis of the reaction gamma p ---> p omega and the search for nucleon resonances,''
  Phys.\ Rev.\ C {\bf 80}, 065209 (2009)
  
\bibitem{Williams:2009ab} 
  M.~Williams {\it et al.} [CLAS],
 % ``Differential cross sections and spin density matrix elements for the reaction gamma p ---> p omega,''
  Phys.\ Rev.\ C {\bf 80}, 065208 (2009)
   

    
\bibitem{Bradford:2006ba} 
  R.~K.~Bradford {\it et al.} [CLAS],
%  ``First measurement of beam-recoil observables C(x) and C(z) in hyperon photoproduction,''
  Phys.\ Rev.\ C {\bf 75}, 035205 (2007)

\bibitem{Bradford:2005pt} 
  R.~Bradford {\it et al.} [CLAS],
%  ``Differential cross sections for gamma + p ---> K+ + Y for Lambda and Sigma0 hyperons,''
  Phys.\ Rev.\ C {\bf 73}, 035202 (2006)

\bibitem{Ho:2017kca} 
  D.~Ho {\it et al.} [CLAS],
%  ``Beam-Target Helicity Asymmetry for $\vec{\gamma} \vec{n} \rightarrow \pi^- p$ in the $N^*$ Resonance Region,''
  Phys.\ Rev.\ Lett.\  {\bf 118}, no. 24, 242002 (2017)
  doi:10.1103/PhysRevLett.118.242002

\bibitem{McCracken:2009ra} 
  M.~E.~McCracken {\it et al.} [CLAS],
%  ``Differential cross section and recoil polarization measurements for the gamma p to K+ Lambda reaction using CLAS at Jefferson Lab,''
  Phys.\ Rev.\ C {\bf 81}, 025201 (2010)    

\bibitem{Paterson:2016vmc} 
  C.~A.~Paterson {\it et al.} [CLAS],
%  ``Photoproduction of $\Lambda$ and $\Sigma^0$ hyperons using linearly polarized photons,''
  Phys.\ Rev.\ C {\bf 93}, no. 6, 065201 (2016)
  doi:10.1103/PhysRevC.93.065201

\bibitem{Strauch:2015zob} 
  S.~Strauch {\it et al.} [CLAS],
%  ``First Measurement of the Polarization Observable E in the $\vec p(\vec \gamma,\pi^+)n$ Reaction up to 2.25 GeV,''
  Phys.\ Lett.\ B {\bf 750}, 53 (2015)
  doi:10.1016/j.physletb.2015.08.053

\bibitem{Dey:2010hh} 
  B.~Dey {\it et al.} [CLAS],
%  ``Differential cross sections and recoil polarizations for the reaction $\gamma p -> K^+ \Sigma^0$, 
  Phys.\ Rev.\ C {\bf 82}, 025202 (2010)
  
\bibitem{McNabb:2003nf} 
  J.~W.~C.~McNabb {\it et al.} [CLAS],
%  ``Hyperon photoproduction in the nucleon resonance region,''
  Phys.\ Rev.\ C {\bf 69}, 042201 (2004)

\bibitem{Golovatch:2018hjk} 
  E.~Golovatch {\it et al.} [CLAS],
%  ``First results on nucleon resonance photocouplings from the $\gamma p \to \pi^+\pi^-p$ reaction,''
  Phys.\ Lett.\ B {\bf 788}, 371 (2019)
 doi:10.1016/j.physletb.2018.10.013

\bibitem{Anisovich:2017pox} 
  A.~V.~Anisovich {\it et al.},
 % ``$N^*\to N \eta^\prime$ decays from photoproduction of $\eta^\prime$-mesons off protons,''
  Phys.\ Lett.\ B {\bf 772}, 247 (2017)
  doi:10.1016/j.physletb.2017.06.052

\bibitem{Compton:2017xkt} 
  N.~Compton {\it et al.} [CLAS],
%  ``Measurement of the differential and total cross sections of the ${\gamma}d{\rightarrow}{K}^{0}\mathrm{{\Lambda}}(p)$ reaction within the resonance region,''
  Phys.\ Rev.\ C {\bf 96}, no. 6, 065201 (2017)
  doi:10.1103/PhysRevC.96.065201

\bibitem{Ho:2018riy} 
  D.~Ho {\it et al.} [CLAS],
% ``Beam-target helicity asymmetry $E$ in $K^{0}\Lambda$ and $K^{0}\Sigma^0$  photoproduction on the neutron,''
  Phys.\ Rev.\ C {\bf 98}, no. 4, 045205 (2018)
  doi:10.1103/PhysRevC.98.045205

 
\bibitem{Anisovich:2011fc} 
  A.~Anisovich, R.~Beck, E.~Klempt, V.~Nikonov, A.~Sarantsev and U.~Thoma,
%  ``Properties of baryon resonances from a multichannel partial wave analysis,''
  Eur.\ Phys.\ J.\ A {\bf 48}, 15 (2012)

  
\bibitem{JuliaDiaz:2007kz} 
  B.~Julia-Diaz, T.-S.~H.~Lee, A.~Matsuyama and T.~Sato,
%  ``Dynamical coupled-channel model of pi N scattering in the W <= 2-GeV nucleon resonance region,''
  Phys.\ Rev.\ C {\bf 76}, 065201 (2007)

\bibitem{Ronchen:2014cna} 
  D.~R\"onchen {\it et al.},
 % ``Photocouplings at the Pole from Pion Photoproduction,''
  Eur.\ Phys.\ J.\ A {\bf 50}, no. 6, 101 (2014)
 
 \bibitem{Anisovich:2017bsk} 
  A.~V.~Anisovich {\it et al.},
 % ``Strong evidence for nucleon resonances near 1900\,MeV,''
  Phys.\ Rev.\ Lett.\  {\bf 119}, no. 6, 062004 (2017)
  doi:10.1103/PhysRevLett.119.062004
  [arXiv:1712.07549 [nucl-ex]].

\bibitem{Edwards:2011jj} 
  R.~G.~Edwards, J.~J.~Dudek, D.~G.~Richards and S.~J.~Wallace, 
%  ``Excited state baryon spectroscopy from lattice QCD,''
  Phys.\ Rev.\ D {\bf 84}, 074508 (2011)
  doi:10.1103/PhysRevD.84.074508
  [arXiv:1104.5152 [hep-ph]].

\bibitem{Chatterjee:2017yhp} 
  S.~Chatterjee, D.~Mishra, B.~Mohanty and S.~Samanta,
 % ``Freezeout systematics due to the hadron spectrum,''
  Phys.\ Rev.\ C {\bf 96}, no. 5, 054907 (2017)
  doi:10.1103/PhysRevC.96.054907
  [arXiv:1708.08152 [nucl-th]].

\bibitem{Anisovich:2015gia} 
  A.~V.~Anisovich {\it et al.},
% ``Evidence for $\Delta(2200)7/2^-$ from photoproduction and consequence for chiral-symmetry restoration at high mass,''%
  Phys.\ Lett.\ B {\bf 766}, 357 (2017)
  doi:10.1016/j.physletb.2016.12.014

\bibitem{Collins:2017vev} 
  P.~Collins {\it et al.} [CLAS],
 % ``Photon beam asymmetry $\Sigma$ in the reaction $\vec{\gamma} p \to p \omega$ for $E_\gamma$ = 1.152 to 1.876 GeV,''
  Phys.\ Lett.\ B {\bf 773}, 112 (2017)
  doi:10.1016/j.physletb.2017.08.015

\bibitem{Akbar:2017uuk} 
  Z.~Akbar {\it et al.} [CLAS],
 % ``Measurement of the helicity asymmetry $E$ in $\omega\to\pi^+\pi^-\pi^0$ photoproduction,''
  Phys.\ Rev.\ C {\bf 96}, no. 6, 065209 (2017)
  doi:10.1103/PhysRevC.96.065209
  [arXiv:1708.02608 [nucl-ex]].

\bibitem{Roy:2017qwv} 
  P.~Roy {\it et al.} [CLAS],
 % ``Measurement of the beam asymmetry $\Sigma$ and the target asymmetry $T$ in the photoproduction of $\omega$ mesons off  
 % the proton using CLAS at Jefferson Laboratory,''
  Phys.\ Rev.\ C {\bf 97}, no. 5, 055202 (2018)
  doi:10.1103/PhysRevC.97.055202

\bibitem{Roy:2018tjs} 
  P.~Roy {\it et al.} [CLAS],
 % ``First Measurements of the Double-Polarization Observables F , P , and H in ? Photoproduction off Transversely Polarized Protons in the N* Resonance Region,''
  Phys.\ Rev.\ Lett.\  {\bf 122}, no. 16, 162301 (2019)
  doi:10.1103/PhysRevLett.122.162301

\bibitem{Anisovich:2017rpe} 
  A.~V.~Anisovich {\it et al.} [CLAS],
%  ``Differential cross sections and polarization observables from CLAS $K$* photoproduction and the search for new $N$* states,''
  Phys.\ Lett.\ B {\bf 771}, 142 (2017).
  doi:10.1016/j.physletb.2017.05.029

\bibitem{Seraydaryan:2013ija} 
  H.~Seraydaryan {\it et al.} [CLAS Collaboration],
%  ``$\phi$-meson photoproduction on Hydrogen in the neutral decay mode,''
  Phys.\ Rev.\ C {\bf 89}, no. 5, 055206 (2014)

\bibitem{Dey:2014tfa} 
  B.~Dey {\it et al.} [CLAS Collaboration],
%  ``Data analysis techniques, differential cross sections, and spin density matrix elements for the reaction $\gamma p \rightarrow \phi p$,''
  Phys.\ Rev.\ C {\bf 89}, no. 5, 055208 (2014)
 
 \bibitem{Lebed:2015dca} 
  R.~F.~Lebed,
 % ``Do the $P_c^+$ pentaquarks have strange siblings?,''
  Phys.\ Rev.\ D {\bf 92}, no. 11, 114030 (2015)

\bibitem{sarantsev2015} A. Sarantsev, talk at this workshop  
 
%\bibitem{Liu:2005pm} 
%  B.~C.~Liu and B.~S.~Zou,
%  %``Mass and K Lambda coupling of N*(1535),''
%  Phys.\ Rev.\ Lett.\  {\bf 96}, 042002 (2006)    

\bibitem{Joo:2001tw} 
  K.~Joo {\it et al.} [CLAS],
 % ``Q**2 dependence of quadrupole strength in the gamma* p ---> Delta+(1232) ---> p pi0 transition,''
  Phys.\ Rev.\ Lett.\  {\bf 88}, 122001 (2002)

\bibitem{Ungaro:2006df} 
  M.~Ungaro {\it et al.} [CLAS],
 % ``Measurement of the N ---> Delta+(1232) transition at high momentum transfer by pi0 electroproduction,''
  Phys.\ Rev.\ Lett.\  {\bf 97}, 112003 (2006)

\bibitem{Frolov:1998pw} 
  V.~V.~Frolov {\it et al.},
%  ``Electroproduction of the Delta (1232) resonance at high momentum transfer,''
  Phys.\ Rev.\ Lett.\  {\bf 82}, 45 (1999)

\bibitem{Aznauryan:2008pe} 
  I.~G.~Aznauryan {\it et al.} [CLAS],
%  ``Electroexcitation of the Roper resonance for 1.7 < Q**2 < 4.5 -GeV2 in vec-ep ---> en pi+,''
  Phys.\ Rev.\ C {\bf 78}, 045209 (2008)

 \bibitem{Aznauryan:2009mx} 
  I.~G.~Aznauryan {\it et al.} [CLAS], 
%  ``Electroexcitation of nucleon resonances from CLAS data on single pion electroproduction,''
  Phys.\ Rev.\ C {\bf 80}, 055203 (2009) 

 \bibitem{Lin:2011da} 
  H.~W.~Lin and S.~D.~Cohen,
%  ``Roper Properties on the Lattice: An Update,''
  AIP Conf.\ Proc.\  {\bf 1432}, no. 1, 305 (2012)
  doi:10.1063/1.3701236
  [arXiv:1108.2528 [hep-lat]].
 
\bibitem{Burkert:2019bhp} 
  V.~D.~Burkert and C.~D.~Roberts,
%  ``Colloquium : Roper resonance: Toward a solution to the fifty year puzzle,''
  Rev.\ Mod.\ Phys.\  {\bf 91}, no. 1, 011003 (2019)
  doi:10.1103/RevModPhys.91.011003
  [arXiv:1710.02549 [nucl-ex]]
  
\bibitem{Mokeev:2012vsa} 
  V.~I.~Mokeev {\it et al.} [CLAS],
%  ``Experimental Study of the $P_{11}(1440)$ and $D_{13}(1520)$ resonances from CLAS data on $ep \rightarrow e'\pi^{+} \pi^{-} p'$,''
  Phys.\ Rev.\ C {\bf 86}, 035203 (2012)

\bibitem{Mokeev:2015lda} 
  V.~I.~Mokeev {\it et al.},
 % ``New Results from the Studies of the $N(1440)1/2^+$, $N(1520)3/2^-$, and $\Delta(1620)1/2^-$ Resonances in Exclusive $ep \to e'p' \pi^+ \pi^-$ Electroproduction with the CLAS Detector,''
  Phys.\ Rev.\ C {\bf 93}, no. 2, 025206 (2016)
  doi:10.1103/PhysRevC.93.025206

\bibitem{Fedotov:2018oan} 
  G.~V.~Fedotov {\it et al.} [CLAS],
 % ``Measurements of the $\gamma_{v} p \rightarrow p' \pi^{+} \pi^{-}$ cross section with the CLAS detector for $0.4$ GeV$^{2}$ $< Q^{2} <$ $1.0$ GeV$^{2}$ and $1.3$ GeV $< W <$ $1.825$ GeV,''
  Phys.\ Rev.\ C {\bf 98}, no. 2, 025203 (2018)
  doi:10.1103/PhysRevC.98.025203

\bibitem{Isupov:2017lnd} 
  E.~L.~Isupov {\it et al.} [CLAS],
%  ``Measurements of $e p \to e' \pi^+ \pi^- p'$ Cross Sections with CLAS at $1.40$ Gev $< W < 2.0$ GeV and $2.0$ GeV$^2$ $< Q^2 < 5.0$ GeV$^2$,''
  Phys.\ Rev.\ C {\bf 96}, no. 2, 025209 (2017)
  doi:10.1103/PhysRevC.96.025209

\bibitem{Denizli:2007tq} 
  H.~Denizli {\it et al.} [CLAS],
 % ``Q*2 dependence of the S(11)(1535) photocoupling and evidence for a P-wave resonance in eta electroproduction,''
  Phys.\ Rev.\ C {\bf 76}, 015204 (2007)

\bibitem{Armstrong:1998wg} 
  C.~S.~Armstrong {\it et al.} [JLab E94014 Collaboration],
 % ``Electroproduction of the S(11)(1535) resonance at high momentum transfer,''
  Phys.\ Rev.\ D {\bf 60}, 052004 (1999)
   
 \bibitem{Egiyan:2006ks} 
  H.~Egiyan {\it et al.} [CLAS],
%  ``Single pi+ electroproduction on the proton in the first and second resonance regions at 0.25-GeV**2 < Q**2 < 0.65-GeV**2 using CLAS,''
  Phys.\ Rev.\ C {\bf 73}, 025204 (2006) 
  
\bibitem{Park:2007tn} 
  K.~Park {\it et al.} [CLAS],
%  ``Cross sections and beam asymmetries for vec(e) p ---> en pi+ in the nucleon resonance region for 1.7 <= Q**2 <= 4.5-(GeV)**2,''
  Phys.\ Rev.\ C {\bf 77}, 015208 (2008)

\bibitem{Park:2014yea} 
  K.~Park {\it et al.} [CLAS],
 % ``Measurements of $ep \to e^\prime \pi^+n$ at W = 1.6 - 2.0 GeV and extraction of nucleon resonance electrocouplings at CLAS,''
  Phys.\ Rev.\ C {\bf 91}, 045203 (2015)

\bibitem{Aznauryan:2015zta} 
  I.~G.~Aznauryan and V.~D.~Burkert,
 % ``Electroexcitation of the $\Delta(1232)\frac{3}{2}^+$ and $\Delta(1600)\frac{3}{2}^+$ in a light-front relativistic quark model,''
  Phys.\ Rev.\ C {\bf 92}, no. 3, 035211 (2015)

\bibitem{Segovia:2015hra} 
  J.~Segovia et al.,   %``Completing the picture of the Roper resonance,''
  Phys.\ Rev.\ Lett.\  {\bf 115}, no. 17, 171801 (2015)

\bibitem{Anikin:2015ita} 
  I.~V.~Anikin, V.~M.~Braun and N.~Offen,
%  ``Electroproduction of the $N^*(1535)$ nucleon resonance in QCD,''
  Phys.\ Rev.\ D {\bf 92}, no. 1, 014018 (2015)

\bibitem{Obukhovsky:2011sc} 
  I.~T.~Obukhovsky et al., 
%  ``Electroproduction of the Roper resonance on the proton: the role of the three-quark core and the molecular $N\sigma$ component,''
  Phys.\ Rev.\ D {\bf 84}, 014004 (2011)
  
\bibitem{Bauer:2014cqa} 
  T.~Bauer, S.~Scherer and L.~Tiator,
%  ``Electromagnetic transition form factors of the Roper resonance in effective field theory,''
  Phys.\ Rev.\ C {\bf 90}, no. 1, 015201 (2014)

\bibitem{Tiator:2008kd} 
  L.~Tiator and M.~Vanderhaeghen,
%  ``Empirical transverse charge densities in the nucleon-to-P(11)(1440) transition,''
  Phys.\ Lett.\ B {\bf 672}, 344 (2009)
  
\bibitem{Aznauryan:2017nkz} 
  I.~G.~Aznauryan and V.~Burkert,
%  ``Electroexcitation of nucleon resonances of the [70,1?] multiplet in a light-front relativistic quark model,''
  Phys.\ Rev.\ C {\bf 95}, no. 6, 065207 (2017)
  doi:10.1103/PhysRevC.95.065207
  [arXiv:1703.01751 [nucl-th]].

\bibitem{Santopinto:2012nq} 
  E.~Santopinto and M.~M.~Giannini,
%  ``Systematic study of longitudinal and transverse helicity amplitudes in the hypercentral constituent quark model,''
  Phys.\ Rev.\ C {\bf 86}, 065202 (2012)
  doi:10.1103/PhysRevC.86.065202
  [arXiv:1506.01207 [nucl-th]].
 
  
   \bibitem{Moorhouse:1966jn} 
  R.~G.~Moorhouse,
%  ``Photoproduction Of N* Resonances In The Quark Model,''
  Phys.\ Rev.\ Lett.\  {\bf 16}, 772 (1966).
  doi:10.1103/PhysRevLett.16.772
 
\bibitem{Li:1991yba} 
  Z.~p.~Li, V.~Burkert and Z.~j.~Li,
%  ``Electroproduction of the Roper resonance as a hybrid state,''
  Phys.\ Rev.\ D {\bf 46}, 70 (1992).

\bibitem{LOI_Hybrids}  
 L. Lanza, talk at this workshop,  A. D' Angelo et al., Jefferson Lab Letter of Intent LOI12-15-004 (2015). 
 
\bibitem{elou19} L.Elouadrhiri, R. Gothe, V. Mokeev,  talks at this workshop.

\bibitem{Burkert:2018nvj} 
  V.~D.~Burkert,
  ``Jefferson Lab at 12 GeV: The Science Program,''
  Ann.\ Rev.\ Nucl.\ Part.\ Sci.\  {\bf 68}, 405 (2018).
  doi:10.1146/annurev-nucl-101917-021129

\bibitem{Burkert:2018bqq} 
  V.~D.~Burkert, L.~Elouadrhiri and F.~X.~Girod,
  %``The pressure distribution inside the proton,''
  Nature {\bf 557}, no. 7705, 396 (2018).
  doi:10.1038/s41586-018-0060-z
  
  \bibitem{Ozdem:2019pkg} 
  U.~\"Ozdem and K.~Azizi,
  %``Gravitational transition form factors of $N(1535) \rightarrow N$,''
  arXiv:1912.06375 [hep-ph].
}
\end{thebibliography}
\end{document}